\documentclass[pre,floatfix,superscriptaddress,twocolumn,showpacs]{revtex4-1}
\usepackage{graphicx}
\usepackage{amsmath,amssymb,amsfonts}
\usepackage{mathtools}
\usepackage{bm}
\usepackage{color}
\usepackage[colorinlistoftodos]{todonotes}
\usepackage{tabularx}

\begin{document}

\title{Classifying the age of a glass based on structural properties: A machine learning approach}
\author{Giulia Janzen}
\affiliation{Department of Applied Physics, Eindhoven University of Technology, P.O.~Box 513, 5600 MB Eindhoven, The Netherlands}
\affiliation{Institute for Complex Molecular Systems, Eindhoven University of Technology, P.O.~Box 513, 5600 MB Eindhoven, The Netherlands}
\author{Casper Smit $^*$}
 
\affiliation{Department of Applied Physics, Eindhoven University of Technology, P.O.~Box 513, 5600 MB Eindhoven, The Netherlands}
\affiliation{Institute of Physics, University of Amsterdam, Science Park 904, Amsterdam, 1098 XH,
The Netherlands}

\author{Samantha Visbeek }
\thanks{These authors contributed equally to this work.}

\affiliation{Department of Applied Physics, Eindhoven University of Technology, P.O.~Box 513, 5600 MB Eindhoven, The Netherlands}
\affiliation{Institute of Physics, University of Amsterdam, Science Park 904, Amsterdam, 1098 XH,
The Netherlands}
\author{Vincent E.~Debets}
\affiliation{Department of Applied Physics, Eindhoven University of Technology, P.O.~Box 513, 5600 MB Eindhoven, The Netherlands}
\affiliation{Institute for Complex Molecular Systems, Eindhoven University of Technology, P.O.~Box 513, 5600 MB Eindhoven, The Netherlands}

\author{Chengjie~Luo}
\affiliation{Department of Applied Physics, Eindhoven University of Technology, P.O.~Box 513, 5600 MB Eindhoven, The Netherlands}
\affiliation{Institute for Complex Molecular Systems, Eindhoven University of Technology, P.O.~Box 513, 5600 MB Eindhoven, The Netherlands}
\author{Cornelis~Storm}
\affiliation{Department of Applied Physics, Eindhoven University of Technology, P.O.~Box 513, 5600 MB Eindhoven, The Netherlands}
\affiliation{Institute for Complex Molecular Systems, Eindhoven University of Technology, P.O.~Box 513, 5600 MB Eindhoven, The Netherlands}
\author{Simone~Ciarella}
\email{simoneciarella@gmail.com}
\affiliation{Department of Applied Physics, Eindhoven University of Technology, P.O.~Box 513, 5600 MB Eindhoven, The Netherlands}
\affiliation{Institute for Complex Molecular Systems, Eindhoven University of Technology, P.O.~Box 513, 5600 MB Eindhoven, The Netherlands}
\affiliation{Laboratoire de Physique de l’Ecole Normale Sup\'erieure, ENS, Universit\'e PSL, CNRS, Sorbonne Universit\'e, Universit\'e de Paris, F-75005 Paris, France}
\affiliation{Netherlands eScience Center, Amsterdam 1098 XG, The Netherlands}
\author{Liesbeth~M.C.~Janssen}
\email{l.m.c.janssen@tue.nl}
\affiliation{Department of Applied Physics, Eindhoven University of Technology, P.O.~Box 513, 5600 MB Eindhoven, The Netherlands}
\affiliation{Institute for Complex Molecular Systems, Eindhoven University of Technology, P.O.~Box 513, 5600 MB Eindhoven, The Netherlands}

\date\today

\begin{abstract}
It is well established that physical aging of amorphous solids is governed by a marked change in dynamical properties as the material becomes older. Conversely, structural properties such as the radial distribution function exhibit only a very weak age dependence, usually deemed negligible with respect to the numerical noise. Here we demonstrate that the extremely weak age-dependent changes in structure are in fact sufficient to reliably assess the age of a glass with the support of machine learning.
We employ a supervised learning method to predict the age of a glass based on the system's instantaneous radial distribution function. Specifically, we train a multilayer perceptron for a model glassformer quenched to different temperatures, and find that this neural network can accurately classify the age of our system across at least four orders of magnitude in time. Our analysis also reveals which structural features encode the most useful information. Overall, this work shows that through the aid of machine learning, a simple structure-dynamics link can indeed be established for physically aged glasses.

\end{abstract}
\maketitle

\section{Introduction}
\begin{figure}
    \centering
    \includegraphics [width=\columnwidth] {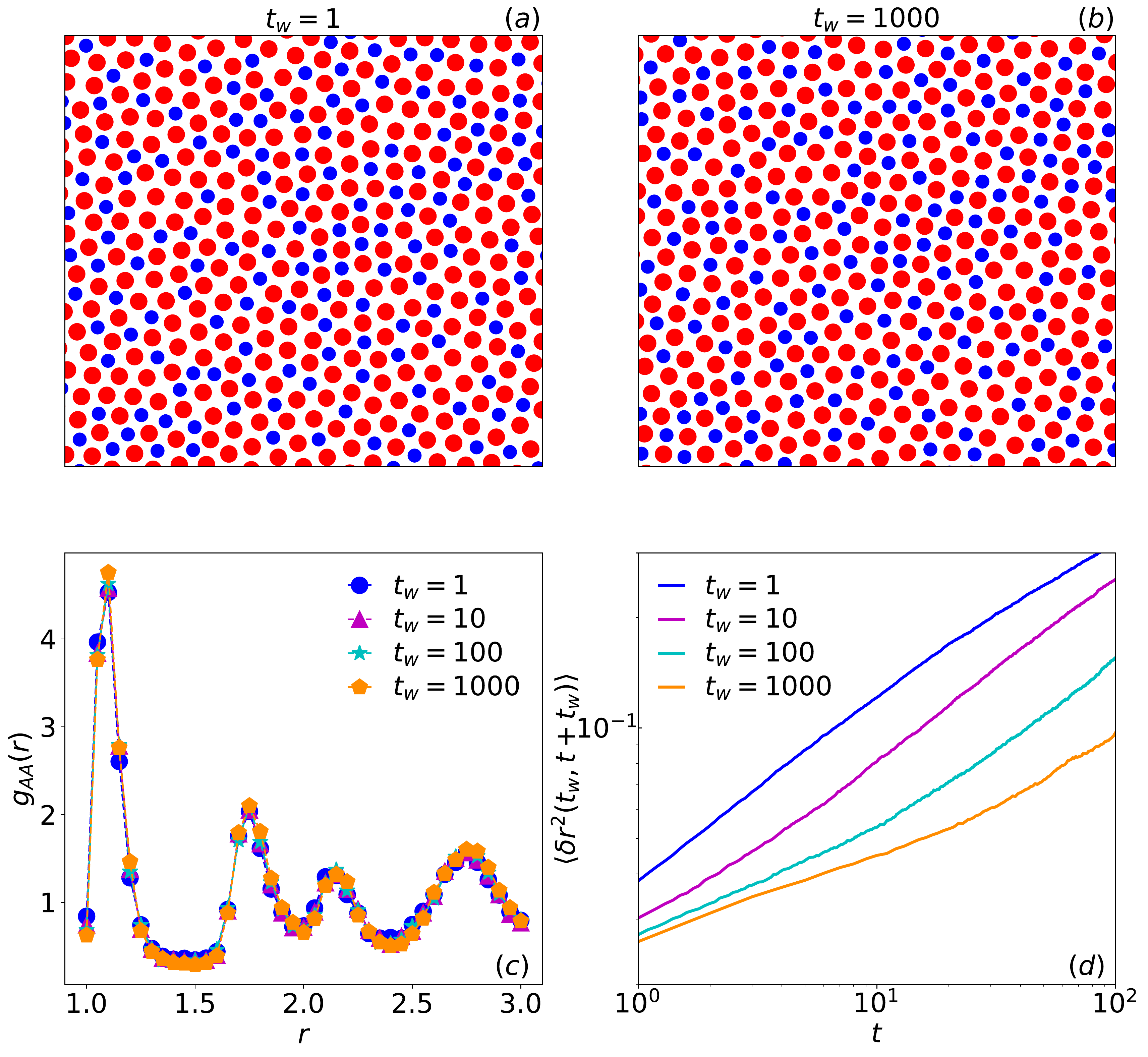} 
   
    \caption{Age dependence in structural and dynamical properties. Snapshots after a temperature quench for waiting time (a) $t_w=1$ and (b) $t_w=1000$. (c) Radial distribution function for A particles $g_{AA}(r)$ and (d) mean-square displacement $\langle \delta r^2 (t_w,t+t_w) \rangle$ at fixed quenching temperature $T_q=0.375$ for waiting times $t_w=1,10,100,1000$. } 
    \label{fig:intro}
    \end{figure}

The structural, dynamical and mechanical properties of a material change as it gets older, i.e.\ it ages~\cite{hodge1995physical,berthier2009statistical,lunkenheimer2005glassy,zhao2013using,Raty2015,Wang2006,odegard2011physical,martin1993aging,mckenna1995evolution}. Physical aging is particularly well studied for glasses due to their slow relaxation dynamics~\cite{struik1977physical,binder2011glassy,biroli2013perspective,debenedetti2001supercooled,PhysRevX.7.031028}. One of the most common methods to study the aging dynamics of a glass consists of a temperature quench toward a lower temperature \cite{kob1997aging,foffi2004aging,Warren_2009,PhysRevLett.110.025501}. After the quench, as the material seeks to recover equilibrium at the new temperature, the relaxation time of the system will increase with its age~\cite{struik1977physical,hutchinson1995physical,barrat1999fluctuation,kob2000fluctuations}. The physical aging in glassy systems can thus be understood as a gradual approach towards increasingly lower-energy equilibrium states~\cite{debenedetti2001supercooled}. It is also well known that, besides a rapid short-time change, the structural properties change only extremely weakly with time~\cite{kob2000aging,PhysRevE.89.062315,PhysRevE.76.031802,Waseda1979,popescu1994structural,PhysRevE.89.062313}. In contrast, the dynamical properties exhibit significant changes over multiple orders of magnitude in time as shown in Fig.\ \ref{fig:intro} and the Supplementary Material \cite{SupplementalMaterial}. It is therefore customary to characterize the aging behavior of a system by means of its dynamical properties.
At the same time, it remains unclear how these strong dynamical changes of an aging glass are connected to its almost constant structure \cite{PhysRevE.76.031802}.

To bridge this gap, Cubuck \textit{et al.} \cite{PhysRevLett.114.108001} have recently developed a pioneering approach which demonstrates that machine learning techniques can in fact successfully correlate structure and dynamics in glassy systems. Cubuck \textit{et al.}\  have introduced a machine learning microscopic structural quantity, so-called softness, which characterizes the local structure around each particle. Based on this approach, several recent works \cite{Schoenholz2016,Cubuk2017,PhysRevE.101.010602,CubukJPC2016,PhysRevMaterials.4.113609,oyama22,tah22,jung22,coslovich22,ciarella22c,Alkemade23,janzen23} have extended our conceptual understanding of glassy liquids by convincingly demonstrating that machine learning is able to accurately connect structural properties with the corresponding dynamics. 
In particular, standard machine learning tools like support vector machines have been able to compute the relaxation time through softness~\cite{Schoenholz263} and collective effects like fragility~\cite{tah22} and low-temperature defects~\cite{ciarella22tls}. More sophisticated models like graph neural networks~\cite{Bapst2020} give accurate predictions of dynamic propensity, but similar results can be achieved by simpler models with accurate structural indicators~\cite{Boattini2021}. 
It is thus evident that machine learning is a powerful tool to study glassy systems and, as suggested by Schoenholz \textit{et al.}\ \cite{Schoenholz263}, it is plausible that it could also be used to shed new light on aging behavior. 

Still, it is a priori not clear whether this level of complexity, both in the machine-learning model and in the input set, is strictly necessary to predict the age of a system from structural properties. 
Indeed, an analysis of the softness suggests that the radial distribution function's first peak contributes the most to predicting rearrangements \cite{Schoenholz2016}. Since the radial distribution function does change weakly with age, one could argue that a traditional approach, which could consist of selecting the radial distribution function’s values that change the most with age and applying linear regression \cite{Bishop2006,mahmoud2019parametric,Tokuda2020}, might already be sufficient to extract the age of a system. However, such a traditional approach would only be expected to work if the uncertainty in the data is sufficiently small, e.g.\ in the thermodynamic limit, while in reality a system is typically finite-sized and thus susceptible to noise.

Here our goal is to classify the age of a glassy system based solely on a snapshot, i.e.\ an instantaneous particle configuration of a finite-sized system. In particular, we compute the radial distribution function at every age and we use this simple feature as input for a neural network. 
We find that a neural network that is trained and tested at a fixed quenching temperature can distinguish between a young and an old glass with $94\%-97\%$ accuracy. We compare our machine learning method with a traditional approach, confirming the superiority of the first. The traditional approach involves manual selection of features that--on average--exhibit the most significant changes with age. However, due to noise, these features yields significantly less reliable age predictions compared to those obtained using features selected through a machine learning approach.
 
In order to analyze our machine learning results in more detail, we perform both a Principal Component Analysis (PCA) and a Shapley Additive Explanation (SHAP). These methods reveal the principal components or the structural features that most strongly encode the age, also allowing us to infer the age of the system from only a subset of the structural data. 
Finally, we explore the role of the quenching temperature, also proving that 
a neural network trained with a set of multiple quenching temperatures generalizes well when tested at a new temperature.
Though we primarily focus on passive systems, we also verify our model for active systems. 
Ultimately, we conclude that a machine learning approach purely based on simple structural properties can reliably infer the age of a glassy system.

    \section{Methods}
\subsection{Simulation model}
We study a two-dimensional (2D) binary mixture of Brownian particles. The overdamped equations of motion for each particle $i$ are given by
\begin{equation}
   \gamma \,\dot{\bm{r}}_i=\sum_{j\neq i } \bm{f}_{ij}+\sqrt{2D_T} \, \bm{\eta} 
   \label{eq:1}
 \end{equation}
where $\bm{r}_i=(x_i,y_i)$ represents the particle's spatial coordinates and the dot denotes the time derivative. The translational diffusion constant is denoted as $D_T=k_B T/\gamma$ and the thermal noise is represented by independent Gaussian stochastic processes $\bm{\eta}=(\eta_x,\eta_y)$  with zero mean and variance $\delta(t-t^\prime)$, where $k_B$ is the Boltzmann constant, $T$ the temperature, and $\gamma$ the friction coefficient. Lastly, $\bm{f}_{ij}=-\nabla_i V(r_{ij})$ is the interaction force between particles $i$ and $j$, where $r_{ij}=\left|\textbf{r}_i-\textbf{r}_j \right|$ and $V$ is a Lennard-Jones potential \cite{Lennard_Jones_1931} with a cutoff distance $r_{ij} = 2.5 \sigma_{ij}$. In order to prevent crystallization we use the 2D binary Kob-Andersen mixture \cite{kob1995testing}: $A=65 \%$, $B=35 \%$,  $\epsilon_{AA}=1$, $\epsilon_{BB}=0.5 \epsilon_{AA} $, $\epsilon_{AB}=1.5 \epsilon_{AA}$, $\sigma_{AA}=1$, $\sigma_{BB}=0.88 \sigma_{AA}$ and $\sigma_{AB}=0.8 \sigma_{AA}$. 
We set the number density to $\rho=1.2$, the number of particles to $N=10\,000$ and $D_T=1$. Results are in reduced units, where $\sigma_{AA}$, $\epsilon_{AA}$, $\frac{\sigma_{AA}^2 \gamma}{\epsilon_{AA}}$, and $\frac{\epsilon_{AA}}{k_B}$ are  the units of length, energy, time, and temperature, respectively. Simulations have been performed using LAMMPS \cite{PLIMPTON19951} by solving Eq.~\ref{eq:1} via the Euler-Maruyama method \cite{kloeden2011numerical} with a step size $\delta t=10^{-4}$.

As additional verification of our method we also study the aging behavior of an active glass. For this we use the active Brownian particle (ABP) model, which combines  thermal motion with a constant self-propulsion speed \cite{romanczuk2012active,ramaswamy2010mechanics,lindner2008diffusion,lowen2020inertial,Shaebani2020,PhysRevX.9.021009}. To obtain the equation of motion for ABPs, in Eq.\ \ref{eq:1} we need to add the self-propulsion term. This term is defined as $f \, \bm{n}_i$, where $f /  \gamma$ is the constant self-propulsion speed along a direction $\bm{n}_i=(\cos{\theta}_i,\sin{\theta}_i)$, $\theta_i$ is the rotational coordinate, and $f$ is the 
magnitude of the active force. The rotational coordinate obeys $\dot{\theta}_i=\sqrt{2D_{r}} \,\eta_{\theta}$, where $D_r$ is the rotational diffusion coefficient and $\eta_{\theta}$ is a Gaussian stochastic process. The persistence time, $\tau_r$, is defined as the inverse of the rotational diffusion coefficient and determines the decay time of a particle's orientation \cite{zottl2016emergent}. Finally, we choose to focus on a relatively large system with $N=10\,000$ particles, but we verified that our machine learning approach also performs well for a smaller system with $N=1\,000$ (Supplementary material \cite{SupplementalMaterial}).

\subsection{Aging}
For our data set, we prepare 20 independent configurations and let each of them equilibrate at the initial temperature $T_i$. In this work we consider $T_i=1$, which corresponds to the liquid phase, but similar results can in principle be obtained for other initial temperatures. Moreover, the dataset consists of 20 independent configurations since these are sufficient to obtain good performance. After the equilibration process we apply a quench to the final temperature $T_q$ that is lower than the glass transition temperature $T_g$ (for this system $T_g \approx 0.4$ \cite{flenner2015fundamental,Li_2016,janzen2021aging}). We use quenching temperatures between $T_q=0.1$ and $T_q=0.375$ and collect data for waiting times between $0$ and $10^4$. It is well known that the relaxation time as a function of the waiting time follows a power law \cite{kob1997aging}. We therefore split the data into 5 different classes following a logarithmic scale, as shown in Tab.\ \ref{tab:class}. Each class consists of $900$ different waiting times $t_w$ that we also refer to as ages, except for class $0$ which consists of only the single age $t_w=0$. In order to have the same amount of data in each class, we save the particle's configurations every $t_{s}$ time units, with $t_{s}$ specified in Table \ref{tab:class}. 

For each age, we compute the radial distribution function $g_i(r)$ averaged over the number of particles, where $i \in \{AA,BB,AB\}$ indicates the interaction pairs. It has been shown that the radial distribution function's first peak is one of the most important features to predict rearrangements \cite{Schoenholz2016}. To verify whether this also applies to the age classification, we compare the results when the radial distribution function includes or excludes the first peak, corresponding to $g_i(r)$ with $\sigma_i \le r<3 \sigma_i$ and $g_i(r')$ with $\sigma_i+0.15 \le r'<3 \sigma_i+0.15$, respectively. In this paper, we will refer to the radial distribution function without the first peak as $\hat{g}_i(r)$. 
To compute $\hat{g}_i(r)$ or $g_i(r)$ we use a bin width of $\delta r=0.05$, resulting in 40 data points for each of the three partial radial distribution functions $g_i(r)$. These 120 structural properties will be used as an input for our machine learning model. The dataset is randomly divided into a training and a test set that includes $70\%$ and $30\%$ of the data, respectively. To verify that this model also works for an active particle system, we study ABPs with an active force $f=0.5$, a persistence time $\tau_r=1$, and a quenching temperature $T_q=0.25$. We chose these parameters such that the relaxation times of the active and passive systems are of the same order of magnitude \cite{janzen2021aging}.

\begin{table}[]
    \centering
    \begin{tabular}{|c|c|c|}
        \hline
        \textbf{class} & \textbf{age} & \boldsymbol{$t_s$} \\
        \hline
        0 & $t_w=0$ &\vphantom{$1^{1^1}$} $10^{-4}$ \\
        \hline
        1 & $10^0\leq t_w<10^1$ &\vphantom{$1^{1^1}$} $10^{-2}$ \\
        \hline
        2 & $10^1\leq t_w<10^2$ &\vphantom{$1^{1^1}$} $10^{-1}$ \\
        \hline
         3 & $10^2\leq t_w<10^3$ &\vphantom{$1^{1^1}$} $10^{0}$ \\
        \hline
        4 & $10^3\leq t_w<10^4$ &\vphantom{$1^{1^1}$} $10^{1}$ \\
        \hline
    \end{tabular}
    \caption{Description of the dataset. Data are collected from age 0 to age $10^4$ and then divided in 5 different classes. The left column shows the label given to each class, the central column the ages that belong to each class, and the right column the time interval $t_s$ used to collect data. }
    \label{tab:class}
\end{table}

\subsection{Classification model}
To carry out the age classification task we use a multilayer perceptron \cite{GARDNER19982627,Multilayer159058} as implemented in Scikit-learn \cite{pedregosa2011scikit}. This neural network (NN) is composed of multiple layers of interconnected neurons. In the first layer, i.e.\ the input layer, the neurons receive the input vector, while the output layer yields the output signals or classifications with an assigned weight. The hidden layers optimize the weights until the neural network's margin of error is minimal \cite{haykin2009neural}. 

In this work we will use two different NN architectures consisting of either four or twelve hidden layers.
In both cases, all hidden layers have 100 nodes except for the last two which have 50 and 30 nodes, respectively. The ADAM algorithm has been used to update the weights \cite{kingma2014adam}.

To evaluate the model we compute the f1-score
\begin{equation*}
    \text{f1-score}=2 \frac{\mathrm{precision} \cdot \mathrm{recall}}{\mathrm{precision}+\mathrm{recall}}
\end{equation*}
where the precision is the sum of true positives across all classes divided by the sum of both true and false positives over all classes, and the recall is the sum of true positives across all classes divided by the sum of true positives and false negatives across all classes. The f1-score reaches its largest value of 1 when the model has perfect precision and recall and its lowest value of 0 if either the precision or the recall is equal to zero. The list of hyperparameters used for the multilayer perceptron is reported in the supplementary material \cite{SupplementalMaterial}.

\subsection{Feature selection: traditional approach, SHAP analysis, and PCA}
\label{sec:feature_selection}
A key aspect of our work is to establish whether machine learning is truly of added value when inferring the age of a finite system, as opposed to a more traditional approach. 
Some signatures of aging in the $g(r)$ have already been observed \cite{Waseda1979,kob2000aging,PhysRevE.89.062315,PhysRevE.76.031802}, and a traditional approach would focus on the features that change the most with age, that usually include the first peak. If the age dependence of these features is linear or polynomial, we could apply a simple algorithm, e.g., linear regression, to make predictions. To verify if this approach is efficient, we compute 
\begin{equation*}
    \langle \delta g_i(t_w,r) \rangle = \Biggl \langle \frac{g_i(t_w,r)-g_i(t_w=1,r)}{g_i(t_w=1,r)}    \Biggr \rangle
\end{equation*}
where $i \in \{AA,BB,AB\}$, $\sigma_i \le r<3 \sigma_i$ and $\langle \dots \rangle$ denotes an average over twenty independent configurations. The variation $\langle \delta g_i(t_w,r) \rangle$ tells us how much the radial distribution function at age $t_w$ changes compared to $g(r)$ obtained for a very young glass, i.e. $t_w=1$. To select the features that are changing the most, we measure $\delta=\max(\langle \delta g_i(t_w,r) \rangle)-\min(\langle \delta g_i(t_w,r) \rangle)$ and select those with the higher $\delta$.

We then compare this traditional approach to our machine learning strategy. The machine-learning model calculates its predictions using all the available data, but we can also identify which features have a stronger influence on the neural network's prediction. We can then verify if these features correspond to those selected with the traditional approach.
Moreover, in order to gain more insight into the machine learning model's prediction, we perform a SHAP analysis \cite{lundberg2017unified} that calculates the relative contribution of each feature to the prediction. Briefly, the SHAP explanation method computes Shapley values incorporating concepts from cooperative game theory.
 The goal of this analysis is to distribute the total payoff among players taking into account the importance of their contribution to the final outcome. In this context, the feature values are the players, the model is the coalition, and the payoff is the model's prediction. 

 Finally, we perform a PCA analysis \cite{jolliffe2002principal}. PCA is a valuable tool for condensing multidimensional data with correlated variables into new variables that represent linear combinations of the original ones. Essentially, PCA serves as a means to reduce the dimensionality of high-dimensional data. Through the identification of variables exhibiting significant variances, we can uncover the inherent characteristics within the data. The first component corresponds to the projection axis that maximizes the variance in a certain direction, while the second principal component is an orthogonal projection axis that maximizes the variance along the next-leading direction. This process can be iterated to identify additional components.

We will explain in the next sections how machine learning approaches outperform the traditional approach for the system under study, demonstrating that machine learning can be more efficient in inferring the age of a material from simple static properties when noise is inherently present in the data.

\section{Results and discussion}
\subsection{Fixed quenching temperature}
Let us first focus on the situation where both the training and prediction have been carried out for a single quenching temperature $T_q$. This allows us to finely tune the machine-learning model. In the following, networks that are trained with a single quenching temperature will be referred to as '$\mathcal{S}$'.
In Sec.~\ref{sec:Tqdep} we compare these models with a more generalized machine-learning model that is trained for a broad range of quenching temperatures.
\begin{table}[]
    \centering
    \begin{tabular}{|c|c|c|c|}
        \hline
      \boldsymbol{$T_q$} \vphantom{$1^{1^1}$}& \textbf{class} & \textbf{f1-score} & \textbf{score} \\
         \hline 
         \multicolumn{4}{|c|}{\textbf{Passive system} } \\
        \hline
           & 0 & 1 &   \\
            & 1 & 0.99 &    \\
          0.1  & 2 & 0.97 & 0.97    \\
            & 3 & 0.96 &    \\
            & 4 & 0.98 &    \\
        \hline
            & 0 & 1 & \\
            & 1 & 0.98 &    \\
          0.25  & 2 & 0.92 &  0.94   \\
            & 3 & 0.91 &    \\
            & 4 & 0.97 &    \\
        \hline
              &0 & 1 &  \\
              & 1 & 0.97 &    \\
          0.375  & 2 & 0.93 & 0.95   \\
              & 3 & 0.94 &    \\
              & 4 & 0.96 &    \\
                \hline
         \multicolumn{4}{|c|}{\textbf{Active system} } \\
          \hline
            & 0 & 1 & \\
            & 1 & 0.96 &    \\
          0.25  & 2 & 0.92 &  0.93   \\
            & 3 & 0.91 &    \\
            & 4 & 0.92 &    \\
    \hline
    \end{tabular}
     
    \caption{Classification performance of the neural networks $\mathcal{S}$ in the passive and active case. The passive neural networks are trained and tested with $T_q=\{0.1,0.25,0.375 \}$, while the active NN is trained and tested with $T_q=0.25$, an active force $f=0.5$, and a persistence time $\tau_r=1$. The model has $g_i(r)$ as input, with $\sigma_i \le r < 3 \sigma_i$. In the left column we show the $T_q$ at which each $\mathcal{S}$ is trained, then we show the class label and its corresponding f1-score, and finally in the last column we provide the overall score obtained in the test set.}
    \label{Tab:singleNN}
    \end{table}
    \label{sec:singleT}
       \begin{figure*}
    \centering
    \includegraphics [width=\textwidth] {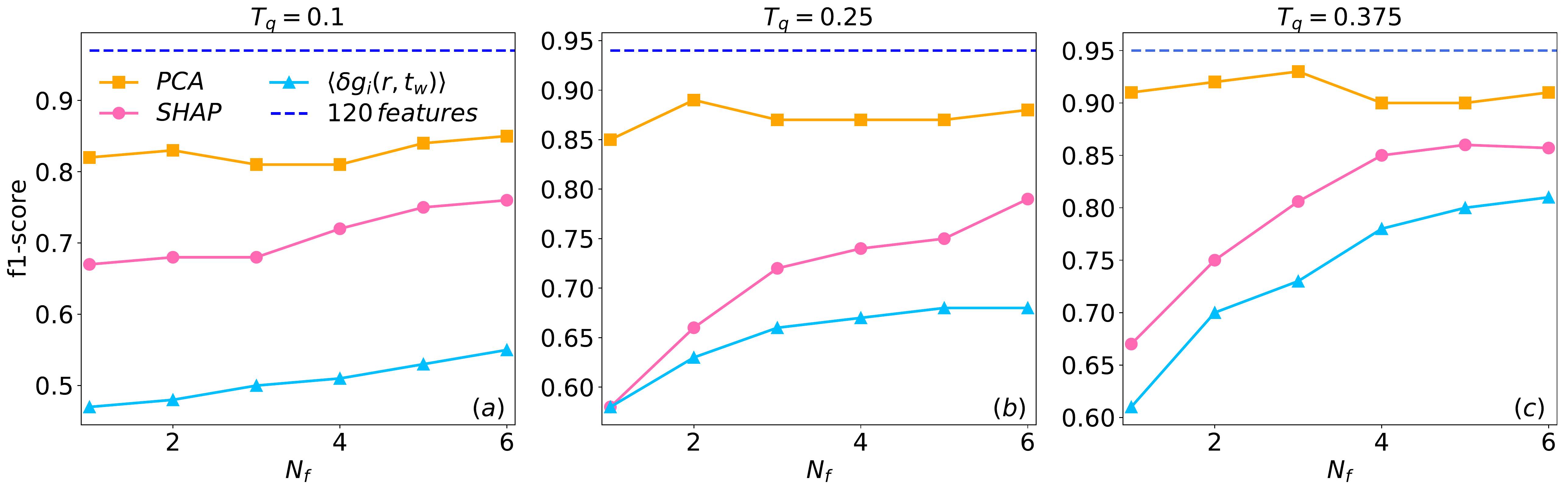} 
    \caption{The f1-score as a function of the number of features $N_f$ used in the machine-learning model. The blue dashed line corresponds to the f1-score showed in Table \ref{Tab:singleNN} obtained with the full dataset (120 features). The orange squares represent the f1-score obtained with the principal components selected by PCA, the pink dots  correspond to the f1-score obtained with the most important features selected with a SHAP analysis, and finally the light blue triangles represent the f1-score gained using the features that change the most on average. (a) Quenching temperature $T_q=0.1$. The  SHAP analysis shows that the  six most important features are: $g_{AA}(1.05)$, $g_{AA}(1.65)$, $g_{AB}(1.35)$, $g_{AA}(1.80)$, $g_{AA}(1.75)$, $g_{AB}(1.30)$. The six features that are changing the most on average are: $g_{AA}(1.00)$, $g_{BB}(2.00)$, $g_{AB}(1.25)$, $g_{BB}(2.05)$, $g_{AA}(1.60)$, $g_{AB}(1.40)$. (b) Quenching temperature $T_q=0.25$. The  SHAP analysis shows that the  six most important features are: $g_{AA}(1.00)$, $g_{AA}(1.65)$, $g_{AB}(1.60)$, $g_{AB}(1.55)$, $g_{BB}(1.80)$, $g_{AA}(1.05)$. The six features that are changing the most on average are: $g_{AA}(1.00)$, $g_{AB}(1.55)$, $g_{BB}(1.00)$, $g_{BB}(2.25)$, $g_{AB}(1.40)$, $g_{BB}(1.05)$.(c) Quenching temperature $T_q=0.375$. The  SHAP analysis shows that the  six most important features are: $g_{BB}(1.50)$, $g_{AA}(1.00)$, $g_{AA}(2.45)$, $g_{AB}(1.60)$, $g_{BB}(1.55)$, $g_{BB}(1.45)$. The six features that are changing the most on average are: $g_{BB}(1.00)$, $g_{BB}(1.55)$, $g_{BB}(1.05)$, $g_{BB}(1.50)$, $g_{BB}(2.35)$, $g_{BB}(2.30)$.}
    \label{fig:score-Nf}
    \end{figure*}

\begin{figure}
    \centering
    \includegraphics [width=\columnwidth] {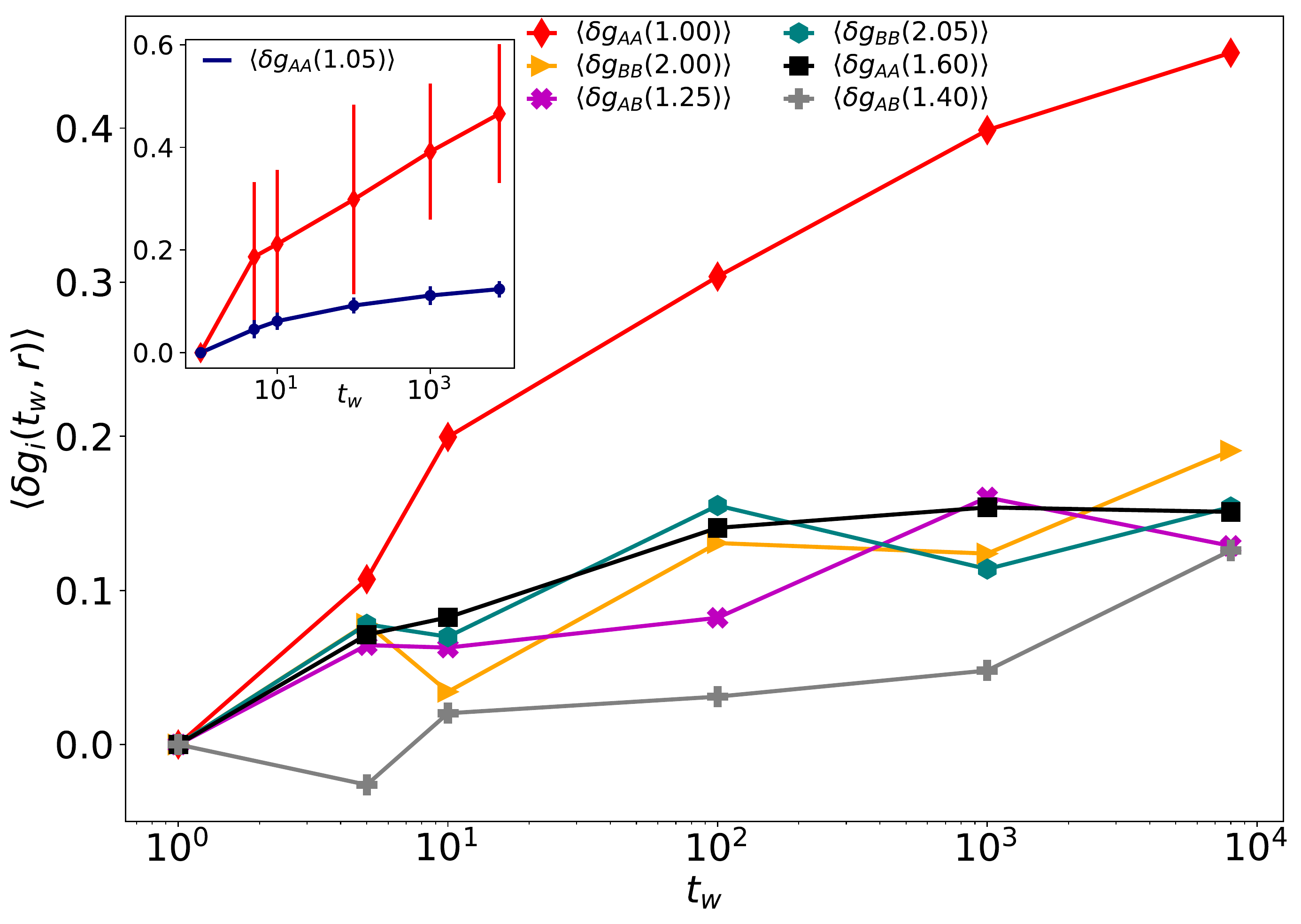} 
    \caption{Variation of the radial distribution function, $\langle \delta g_i(t_w,r) \rangle$, as a function of age $t_w$. The symbols represent the six features that change the most on average for a system at $T_q=0.1$. In the inset the red diamonds show the feature that changes the most, $\langle \delta g_{AA}(1.00) \rangle$, with its corresponding standard deviation, and the blue dots correspond to the most important feature according to the SHAP analysis, $\langle \delta g_{AA}(1.05) \rangle$, and its standard deviation.}
    \label{fig:deltag}
    \end{figure}
\subsubsection{Age prediction}
To infer the age of a glass, we use a NN that only uses the instantaneous radial distribution functions $g_i(r)$ (120 features in total) as input. 
We train three different neural networks $\mathcal{S}$, composed by four hidden layers, and trained and tested at quenching temperatures $T_q=0.1,0.25$ and $0.375$, respectively. We have verified that our bigger alternative NN with twelve hidden layers does not improve the performances (see supplementary material \cite{SupplementalMaterial}). In Table \ref{Tab:singleNN}, we show the f1-score for each class and the overall score computed in the test set. Table \ref{Tab:singleNN} shows that the f1-score for each class is always higher than $0.9$, regardless of the quenching temperature. From this excellent score across all age categories it is clear that, even if the waiting time dependence of the radial distribution function is considered weak, a NN trained exclusively on this structural property is able to distinguish between a young and an old glass with remarkable accuracy. 

To verify whether our machine learning approach can also classify the age of an active system, we train and test a model $\mathcal{S}$ with the data of dense ABPs. In Tab.~\ref{Tab:singleNN} we show the f1-score corresponding to an aging active system ($f=0.5$, $\tau_r=1$) at quenching temperature $T_q=0.25$. As in the passive case, the f1-scores exceed 0.9 for all age categories across four decades in time. Thus, the neural network also performs well for active glasses when trained and tested at the same temperature. This is consistent with recent works \cite{mandal2020multiple,janzen2021aging} demonstrating that an active system's aging behavior shares several similarities with a passive glass, notably the power-law growth of the alpha relaxation time as a function of the waiting time. In particular, this explains why our machine-learning models for passive and active systems have a similar predictive performance. Finally, while we focused on a two-dimensional system, we have verified that this model also works for a three-dimensional system \cite{SupplementalMaterial}.

\subsubsection{Traditional approach versus machine learning}
\label{Sec:Experimental}
In the previous section we have shown that a NN trained with $120$ static features can reliably predict the age of the system at a given temperature. Here we explore whether all these features are necessary to train a well-performing model, since a subset of features might already efficiently encode the age of the material. To this end, we sort all $g_i(r)$ features in order of importance; The order is determined either from a traditional approach that simply looks for the values of $g_i(r)$ changing the most with age, a machine-learning-based SHAP analysis, or a PCA analysis which extracts the most important components (see Sec.\ \ref{sec:feature_selection}). For these three sortings, we can then train a NN with only the most important features and establish how the age can be most efficiently predicted from minimal structural information.

To compare the traditional approach with machine learning, we train neural networks $\mathcal{S}$ with a different number of features $N_f$, where $N_f \in \{1,2,3,4,5,6\}$. In Fig.\ \ref{fig:score-Nf} we show the f1-score as a function of $N_f$ for both the traditional and machine techniques, namely SHAP-based feature selection and PCA. Each panel corresponds to a different quenching temperature. 
It can be seen that for all considered temperatures, the predictions restricted to the features selected by SHAP are better than the traditional approach, demonstrating that machine-learning-based feature selection is superior to the traditional 'human learning' approach in this case. Furthermore, PCA outperforms both the conventional method and the SHAP analysis. Importantly, however, the variance associated with the number of principal components ranging from one to six is below 0.5, meaning that the first six components do not contain all the information within the dataset (see Supplementary Material \cite{SupplementalMaterial}). Consequently, this suggests that the dimensionality-reduced dataset from PCA does not comprehensively represent the entire dataset. 

Even though the f1-score for a restricted model is always lower than that for the full model with 120 features (see Fig.\ \ref{fig:score-Nf}), both PCA and SHAP restricted to $N_f=6$ can be considered good classifiers, since their f1-scores are always greater than $0.76$ and $0.80$, respectively. However, the list of the six optimal features with both SHAP and PCA changes with the quenching temperature, while the full model leads to a f1-score higher than $0.9$ regardless of the quenching temperature.
Thus, while fewer features can indeed be used to obtain good predictions, this comes at the price of performing a new PCA or SHAP analysis for the full model at each temperature, and hence the full model can be deemed more efficient overall. 

Let us now inspect the feature selection more closely to determine why the machine-learning-based selection outperforms the traditional approach. To compare the most important features selected by these two approaches, the focus of the remainder of this section will be on the SHAP analysis and the traditional approach. We note that one could also perform a more in-depth analysis of the main PCA components, but due to their relatively small variance, we prefer to focus on SHAP instead. 

A key point in support of machine learning is its ability to perform well for noisy data, i.e.\ in the presence of fluctuations that are 
inevitable in experimental or simulation data of finite-sized systems. Figure \ref{fig:deltag} shows the six features that change the most \textit{on average} for a system with quenching temperature $T_q=0.1$. From this plot it is clear that $g_{AA}(1.00)$ is the feature that varies the most with age. In particular, an older system corresponds to a larger value of $\langle \delta g_{AA}(1.00) \rangle$. Therefore, one could argue that this feature alone should be sufficient to predict the age of the system. However, Fig.\ \ref{fig:score-Nf}(a) shows that the f1-score obtained from a NN trained with only $g_{AA}(1.00)$ (light blue point at $N_f=1$) is lower than $0.5$, while the one corresponding to a single SHAP-selected feature (pink point at $N_f=1$) is greater than $0.6$. This single most important feature according to SHAP is $g_{AA}(1.05)$, which the machine-learning model selects even if it does not change much with age (inset Fig. \ref{fig:deltag}). The reason for this choice, also highlighted in the inset of Fig. \ref{fig:deltag}, is that the \textit{standard deviation} associated to $\langle \delta g_{AA}(1.00) \rangle$ is much larger than the one obtained for $\langle \delta g_{AA}(1.05) \rangle$. 
Furthermore, in the supplementary material \cite{SupplementalMaterial}, we present the results achieved by subtracting the mean radial distribution function of the complete dataset from the radial distribution function at different ages, denoted as $g_{i}(r)-\langle g_{i}(r) \rangle$ with $i \in {AA,BB,AB}$, at $T_q=0.375$ for a single snapshot. This analysis confirms results similar to those reported in Fig.~\ref{fig:deltag}, and we do not observe any additional significant effects.

From this analysis we can conclude that, according to SHAP, the features that have the biggest influence on the model's prediction are not necessarily those that change the most with age, but rather features that change monotonically with age and have a relatively small standard deviation. Overall, we see that the noise associated with an instantaneous configuration is usually too large to make reliable age predictions based on features selected with the traditional approach.
Therefore, we conclude that in order to properly classify the age of a glass from a single snapshot, a machine learning approach is preferred, since it is better equipped to handle noise.

\subsection{Quenching temperature dependence}
\label{sec:Tqdep}

\subsubsection{Age prediction with a generalized model}
\label{generalized model}
 \begin{figure}
    \centering
    \includegraphics [width=\columnwidth] {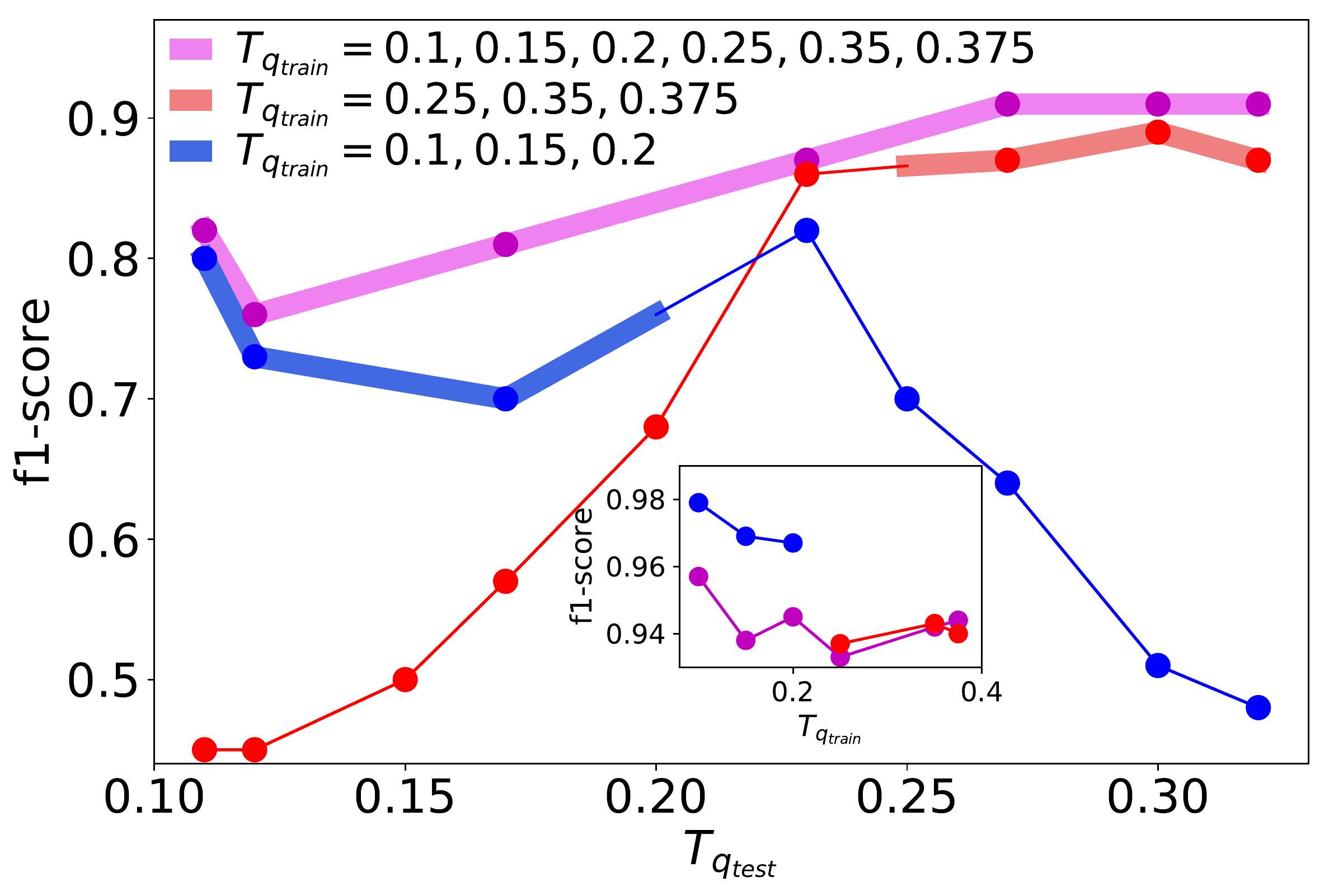} 
    \caption{The f1-score as a function of the quenching temperature used to test the model $T_{q_{test}}$. The purple line shows the neural network $\mathcal{M}$ trained with $T_{q_{train}}=0.1,0.15,0.2,0.25,0.35,0.375$, the red line represents $\mathcal{M}_{high}$ trained with $T_{q_{train}}=0.25,0.35,$ $0.375$ and the blue line indicates $\mathcal{M}_{low}$ trained for $T_{q_{train}}=0.1,0.15,0.2$. Each dot corresponds to the f1-score obtained in the test set when $T_{q_{test}} \ne T_{q_{train}}$. The temperatures used in the test set are $T_{q_{test}}=0.1,0.11,0.12,0.15,0.17,0.2,0.23,0.25,0.27,0.3,0.32$. The inset shows the f1-score when the networks $\mathcal{M}$, $\mathcal{M}_{high}$ and $\mathcal{M}_{low}$ are tested with $T_{q_{test}}=T_{q_{train}}$.}
    \label{fig:score}
    \end{figure}

We now aim to build a general model that is able to classify the age of the system at any quenching temperature regardless of the $T_q$ used in the training. 
The first attempt to achieve this goal consists of determining whether the model $\mathcal{S}$, introduced in the previous section \ref{sec:singleT}, can correctly classify unseen data at different temperatures. Therefore, we test each neural network $\mathcal{S}$ with $T_{q_{test}}=0.11,0.12,0.15,0.17,0.2,0.23,0.3,0.32,0.35$.
Our results show that the model $\mathcal{S}$ trained with the partial radial distribution functions without the first peaks, $\hat{g}_i(r)$, generalizes better than when trained with the full radial distribution functions $g_i(r)$ \cite{SupplementalMaterial}.
This is not only due to the strong temperature dependence of the main peaks, but also to the fact that those data points are extremely noisy (as shown in Sec.\ \ref{Sec:Experimental}). For this reason, in this section we will focus on the results corresponding to neural networks trained with $\hat{g}_i(r)$. Moreover, we have found that this model can extrapolate reasonably well only when the difference between $T_{q_{train}}$ and $T_{q_{test}}$ remains sufficiently small. Since this model cannot be used to predict the age of the system at an arbitrary quenching temperature, we examine whether the performance of our model further improves when it is trained with a set of multiple (in our case three or six) different quenching temperatures.

To this end we use a new neural network with twelve hidden layers, referred to as '$\mathcal{M}$', that is trained with $T_{q_{train}}=0.1,0.15,0.2,0.25,0.35,0.375$, and subsequently tested with $T_{q_{test}}=0.11,0.12,0.17,0.23,0.27,0.3,0.32$. 
We have also verified that for this dataset a NN with twelve hidden layers generalizes better compared to a smaller network (see supplementary material \cite{SupplementalMaterial}), and that using $T_q$ and $\hat{g}_i(r)$ as input yields the best performance. 
The purple line in Fig.~\ref{fig:score} shows the f1-score of our most general model $\mathcal{M}$ as a function of $T_{q_{test}}$. It can be seen that the f1-score in the test set is always higher than $0.76$. Therefore, this model is able to interpolate reasonably well for unseen data. Specifically, for $0.11 \le T_{q_{test}} \le 0.17$ we find that $0.76 \le$ f1-score $\le 0.82$, while, when $0.23 \le T_{q_{test}} \le 0.32$ the model has $0.87 \le$ f1-score $\le 0.91$. Our neural network $\mathcal{M}$ thus performs better for the higher quenching temperatures, i.e.\ when $T_{q_{test}} \ge 0.23$.

To better understand this behavior and to test if different aging regimes exist, we split the training set into two parts: One for low temperatures $\mathcal{M}_{low}$, with $T_{q_{train}}=0.1,0.15,0.2$, and one for higher quenching temperatures $\mathcal{M}_{high}$, with $T_{q_{train}}=0.25,0.35,0.375$. In both cases we use a NN with four hidden layers, because for these two datasets this performs better than a larger NN. From Fig. \ref{fig:score} we can see that $\mathcal{M}_{high}$ (red curve) performs well (f1-score $\ge 0.86$) for $T_{q_{test}} \ge 0.23$. For these $T_{q_{test}}$ values the f1-score is very similar to the one obtained with $\mathcal{M}$. This means that for high temperatures even a small network trained with a smaller set of quenching temperatures is able to generalize to quenching temperatures close to those used in the training set. However, when $\mathcal{M}_{high}$ is tested with $T_{q_{test}}<0.23$ the corresponding f1-score is lower than $0.7$. For these temperatures $\mathcal{M}_{high}$ systematically overestimates the age of the system (see supplementary material \cite{SupplementalMaterial}). 
For lower quenching temperatures, instead, $\mathcal{M}_{low}$ (blue curve) has a f1-score higher than $0.7$ when it is tested with $T_{q_{test}} \le 0.25$. In this case this NN performs worse compared to $\mathcal{M}$, but $\mathcal{M}_{low}$ is able to generalize in a larger range of $T_{q_{test}}$ compared to $\mathcal{S}$ trained with $T_{q_{train}}=0.1$ (reported in the Supplementary material \cite{SupplementalMaterial}). In order to have higher performances, the low temperature regime needs a bigger set of $T_{q_{train}}$ and a bigger NN. Moreover, similarly to $\mathcal{M}_{high}$, $\mathcal{M}_{low}$ has a f1-score lower than $0.7$ when tested with $T_{q_{test}}>0.23$. In this case, $\mathcal{M}_{low}$ underestimates the age of the system (see supplementary material \cite{SupplementalMaterial}).
As we shall discuss in the following section, the over- or underestimation of $\mathcal{M}_{high}$ and $\mathcal{M}_{low}$ in unseen temperature ranges might be related to the true underlying physics, as the rate of aging depends on the quenching temperature.

The inset of Fig. \ref{fig:score} shows that when we test the three neural networks ($\mathcal{M}$, $\mathcal{M}_{high}$, and $\mathcal{M}_{low}$) with $T_{q_{test}}=T_{q_{train}}$, the f1-score is always higher than $0.93$, i.e., all models yield excellent predictions when tested for the temperatures they were trained for. From Fig.\ \ref{fig:score}, we can conclude that a NN trained with quenching temperatures $T_{q_{train}}=T_{q_1}, \dots, T_{q_m}$ performs well when tested with $T_{q_1} \le T_{q_{test}} \le T_{q_m}$. Schoenholtz \textit{et al.} \cite{Schoenholz263} have shown that the history-dependent dynamics in glassy systems can be quantified by the softness and that this property can be used to predict $t_w$ even for systems at different temperatures. Our results show that a simpler model, based only on the radial distribution function, can predict the age of a system at any temperature if the NN is trained on a set of multiple quenching temperatures.

Finally, we have also verified that the model $\mathcal{M}$ trained with passive data can correctly classify the age of an unseen active system ($f=0.5$ and $T_q=0.25$) with an f1-score equal to $0.85$. This remarkably good performance can be rationalized as follows. In the steady state, an active system can be mapped onto a passive system using an effective temperature, while during aging the effective temperature will change with the age of the system \cite{janzen2021aging}. In this context each class will correspond to a different effective temperature, and for this reason the NN trained on a passive system with multiple quenching temperatures has high performances when tested on an active system. 
\begin{figure}
    \centering
    \includegraphics [width=\columnwidth] {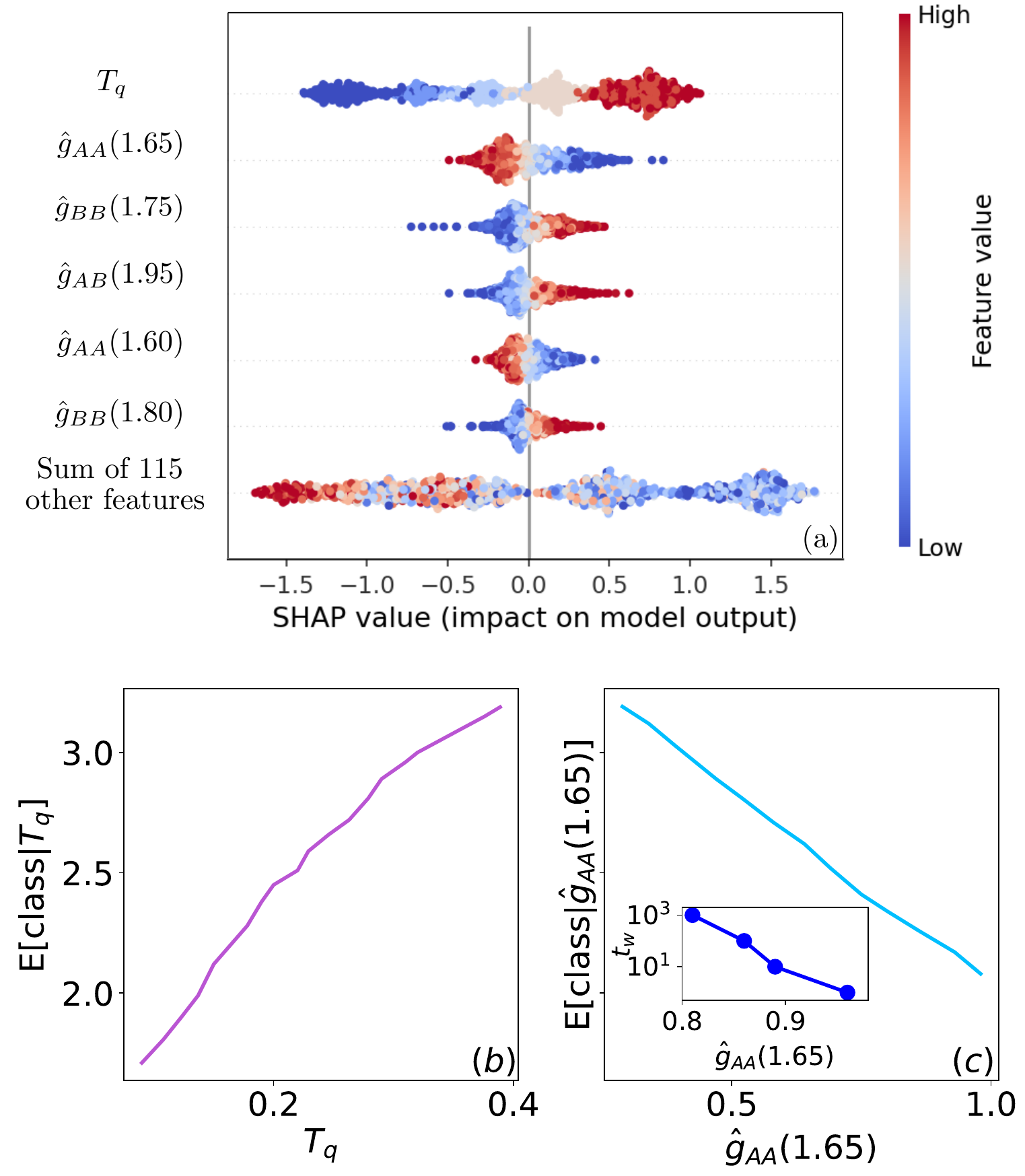} 

    \caption{SHAP-based interpretation of the multilayer perceptron predictions. Here we analyze the neural network $\mathcal{M}$ with twelve hidden layers that has $T_q$ and $\hat{g}_i(r)$ as input and is trained with $T_{q_{train}}=0.1,0.15,0.2,0.25,0.35,0.375$. 
    (a) SHAP beeswarm plot that shows how the most important features impact the model's output. The $x$ position of the dots is determined by the SHAP values of the features and color is used to display the original value of the features. Partial dependence plot for (b) $T_q$ and (c) $\hat g_{AA}(1.65)$. The $x$-axis is the value of the feature and the $y$-axis is the average value of the model output when we fix $T_q$ or $\hat g_{AA}(1.65)$ to a given value. Each class has a label that goes from $0$, young glass with $t_w=0$, to $4$, old glass with $10^3 \le t_w \le 10^4$. The inset of panel (c) shows the waiting time $t_w$ as a function of $\hat g_{AA}(1.65)$. Here we show the actual data for a passive system quenched at $T_q=0.35$.}
    \label{fig:shap}
    \end{figure}
\subsubsection{Physical interpretation of the most important features}
Lastly, we aim to identify which features have a bigger impact on model $\mathcal{M}$'s predictions and how to interpret the machine learning approach from a physical point of view. For this identification, we choose to employ a SHAP analysis, but it is important to mention that a comparable analysis could also be carried out using methods like PCA. The results of this analysis for our most general model trained on all quenching temperatures (purple line in Fig.\ \ref{fig:score}) are presented in Fig.\ \ref{fig:shap}.
 In particular, Fig.\ \ref{fig:shap}(a) shows the SHAP beeswarm plots which indicate the six most important features and how the values of these features influence the model's predictions. The quenching temperature $T_q$ is seen to be the most important feature and the colors in Figure \ref{fig:shap}(a) show that the model interprets low values of $T_q$ as a young glass and high values of $T_q$ as an old glass. 
 
  To better understand this behavior, we have also plotted the partial dependence of $T_q$ in Fig.\ \ref{fig:shap}(b). In this plot the quenching temperature is handled independently from the other features, allowing us to precisely pinpoint how changing $T_q$ impacts the model's predictions. In agreement with Fig.\ \ref{fig:shap}(a), this plot shows that according to the NN a low $T_q$ is more likely to correspond to a young glass. At first glance this interpretation may look incorrect since the dataset consists of the same amount of ages for each temperature. However, at any fixed waiting time $t_w$, a system quenched to a higher $T_q$ is always closer to its steady state  compared to a system at a lower quenching temperature, because its temperature jump is smaller. Therefore, for any given $t_w$, the system at a higher $T_q$ is effectively older than the one quenched to a lower $T_q$. This analysis shows that the NN understands that glasses quenched at higher temperatures age faster. Therefore, the misclassification of $\mathcal{M}_{high}$ and $\mathcal{M}_{low}$ at low and high temperatures (as shown in Sec.\ \ref{generalized model}), respectively, might be due to the model's ability to learn that the rate of aging depends on the quenching temperature.

Finally, let us look at the most important structural feature for model $\mathcal{M}$'s predictions. 
In Fig.\ \ref{fig:shap}(a) it is shown that the most important structural feature is $\hat g_{AA}(1.65)$, i.e., the point just before the second peak of $\hat g_{AA}$. As discussed in Sec.\ \ref{generalized model}, the main peak of the radial distribution function strongly depends on temperature and is affected by noise. Therefore, we excluded the first peak from the dataset. Our work does not necessarily imply that the main peak is unimportant, and indeed Schoenholz \textit{et al.}\ \cite{Schoenholz2016} have shown that the radial distribution function's first peak gives 77\% accuracy to predict rearrangements. Rather, our work shows that even without the main peak, and focusing only on a seemingly small feature as $\hat g_{AA}(1.65)$, we can reliably classify the age. Thus, even a region where the correlation between particles is low contains enough information to classify the system's age. Moreover, in  Fig.\ \ref{fig:shap}(c) we show that the NN interprets large values of $\hat g_{AA}(1.65)$ as an old glass. This feature interpretation is in agreement with the data, as shown in the inset of Fig.\ \ref{fig:shap}(c).

\section{Conclusions}

In summary, this proof-of-principle study demonstrates that a simple supervised machine learning method can accurately classify the age of a glass undergoing a temperature quench, relying only on partial radial distribution functions (obtained from an instantaneous configuration, averaged over all particles). The performance of our machine learning algorithm is extremely accurate when the quenching temperature $T_q$ used during training is equal to the one used in the test set (model $\mathcal{S}$), and the model also generalizes well to datasets consisting of multiple quenching temperatures (model $\mathcal{M}$). This good performance for various temperatures indicates the robustness of our method. Extrapolation to unseen temperatures outside the training window is also reasonable, provided that the temperature difference is not too large. When extrapolating to significantly lower or higher temperatures, however, we find that our neural network tends to systematically under- or overestimate the age of the glass, respectively.
This breakdown of the model extrapolation could be ultimately driven by a different physical behavior, as it is well known that higher-temperature glasses effectively age faster.

To establish which features in the radial distribution functions best encode the age of a glassy configuration, we have compared a traditional approach based on physical intuition with a machine learning-based analysis employing SHAP or PCA. The traditional approach manually seeks the values of the radial distribution functions that--on average--change the most with age, while the SHAP method extracts the most important features from a trained neural network. This comparison reveals that machine learning methods strongly outperform the more traditional one.
The reason for this is the inevitable statistical noise in the data. Indeed, the fluctuations in the radial distribution functions can vary significantly among different configurations, and the machine-learning model is able to adequately filter out these statistical fluctuations. However, the list of key features selected by SHAP or the principal components selected by PCA changes with the quenching temperature (see Supplementary Material \cite{SupplementalMaterial}). 
It follows that in order to identify the most important structural features, one should in principle train a neural network at each $T_q$ with the full dataset and later perform a SHAP analysis or PCA to identify the key features. Since there is usually no cost associated to using a larger number of features, overall we conclude that a model trained with the full data set (120 features) is the most efficient approach.

For our most general machine-learning model (model $\mathcal{M}$), we have also employed SHAP to explain the predictions. This analysis shows that the two most important features are the quenching temperature $T_q$ and the partial radial distribution function $\hat g_{AA}(1.65)$. Interestingly, the model is thus able to learn that the rate of aging depends on the quenching temperature and, surprisingly, that $\hat g_{AA}(1.65)$, the point just before the radial distribution function's second peak, contains enough information to predict the system's age.

While we have focused on the age classification of a passive glass, we have verified that this machine-learning model works remarkably well even for an active glass composed of active Brownian particles. Our results show that model $\mathcal{M}$ trained with passive data can correctly classify the age of an active system. Therefore, this method could also be used to map the aging behavior of an active glass onto a passive glass at different quenching temperatures \cite{janzen2021aging}.

A potential next step of this work could involve incorporating additional structural descriptors to further investigate the relationship between structure and dynamics in aged glasses. Since in recent years, Smooth Overlap of Atomic Positions (SOAP) parameters have proven to be effective in encoding atomic structures \cite{bartok13,C6CP00415F,coslovich22}, one could explore training a machine learning algorithm using these parameters as input.

Our work demonstrates that, even though the radial distribution function of an aging glass is usually considered to remain constant with age, the age dependence, albeit subtle, is already fully encoded in this simple structural property. We thus argue that machine learning methods can be of true added value compared to traditional physical approaches, since they can uncover previously unseen correlations that would be difficult if not impossible to detect with the human eye. Owing to the simplicity and computational efficiency of our approach, we envision that our machine-learning method can be used in a variety of applications, e.g. to quickly distinguish a system that has already reached its steady state from a system that is still aging. This could be particularly attractive for studies in which physical aging is an undesirable and difficult problem, such as equilibration of deeply supercooled liquids; With our model, it would be possible to verify whether a supercooled liquid has reached equilibrium from a single snapshot.

\begin{acknowledgements}
It is a pleasure to thank Robert Jack for stimulating discussions. This work has been financially supported by the Dutch Research Council (NWO) through a START-UP grant (VED, CL, and LMCJ), Physics Projectruimte grant (GJ and LMCJ), and Vidi grant (LMCJ).
\end{acknowledgements}

 \bibliographystyle{apsrev4-1} 
\bibliography{./aging}

\begin{thebibliography}{72}%
\makeatletter
\providecommand \@ifxundefined [1]{%
 \@ifx{#1\undefined}
}%
\providecommand \@ifnum [1]{%
 \ifnum #1\expandafter \@firstoftwo
 \else \expandafter \@secondoftwo
 \fi
}%
\providecommand \@ifx [1]{%
 \ifx #1\expandafter \@firstoftwo
 \else \expandafter \@secondoftwo
 \fi
}%
\providecommand \natexlab [1]{#1}%
\providecommand \enquote  [1]{``#1''}%
\providecommand \bibnamefont  [1]{#1}%
\providecommand \bibfnamefont [1]{#1}%
\providecommand \citenamefont [1]{#1}%
\providecommand \href@noop [0]{\@secondoftwo}%
\providecommand \href [0]{\begingroup \@sanitize@url \@href}%
\providecommand \@href[1]{\@@startlink{#1}\@@href}%
\providecommand \@@href[1]{\endgroup#1\@@endlink}%
\providecommand \@sanitize@url [0]{\catcode `\\12\catcode `\$12\catcode
  `\&12\catcode `\#12\catcode `\^12\catcode `\_12\catcode `\%12\relax}%
\providecommand \@@startlink[1]{}%
\providecommand \@@endlink[0]{}%
\providecommand \url  [0]{\begingroup\@sanitize@url \@url }%
\providecommand \@url [1]{\endgroup\@href {#1}{\urlprefix }}%
\providecommand \urlprefix  [0]{URL }%
\providecommand \Eprint [0]{\href }%
\providecommand \doibase [0]{http://dx.doi.org/}%
\providecommand \selectlanguage [0]{\@gobble}%
\providecommand \bibinfo  [0]{\@secondoftwo}%
\providecommand \bibfield  [0]{\@secondoftwo}%
\providecommand \translation [1]{[#1]}%
\providecommand \BibitemOpen [0]{}%
\providecommand \bibitemStop [0]{}%
\providecommand \bibitemNoStop [0]{.\EOS\space}%
\providecommand \EOS [0]{\spacefactor3000\relax}%
\providecommand \BibitemShut  [1]{\csname bibitem#1\endcsname}%
\let\auto@bib@innerbib\@empty
\bibitem [{\citenamefont {Hodge}(1995)}]{hodge1995physical}%
  \BibitemOpen
  \bibfield  {author} {\bibinfo {author} {\bibfnamefont {I.~M.}\ \bibnamefont
  {Hodge}},\ }\href@noop {} {\bibfield  {journal} {\bibinfo  {journal}
  {Science}\ }\textbf {\bibinfo {volume} {267}},\ \bibinfo {pages} {1945}
  (\bibinfo {year} {1995})}\BibitemShut {NoStop}%
\bibitem [{\citenamefont {Berthier}\ and\ \citenamefont
  {Biroli}(2009)}]{berthier2009statistical}%
  \BibitemOpen
  \bibfield  {author} {\bibinfo {author} {\bibfnamefont {L.}~\bibnamefont
  {Berthier}}\ and\ \bibinfo {author} {\bibfnamefont {G.}~\bibnamefont
  {Biroli}},\ }\href@noop {} {\bibfield  {journal} {\bibinfo  {journal}
  {Encyclopedia of Complexity and Systems Science}\ ,\ \bibinfo {pages} {4209}}
  (\bibinfo {year} {2009})}\BibitemShut {NoStop}%
\bibitem [{\citenamefont {Lunkenheimer}\ \emph {et~al.}(2005)\citenamefont
  {Lunkenheimer}, \citenamefont {Wehn}, \citenamefont {Schneider},\ and\
  \citenamefont {Loidl}}]{lunkenheimer2005glassy}%
  \BibitemOpen
  \bibfield  {author} {\bibinfo {author} {\bibfnamefont {P.}~\bibnamefont
  {Lunkenheimer}}, \bibinfo {author} {\bibfnamefont {R.}~\bibnamefont {Wehn}},
  \bibinfo {author} {\bibfnamefont {U.}~\bibnamefont {Schneider}}, \ and\
  \bibinfo {author} {\bibfnamefont {A.}~\bibnamefont {Loidl}},\ }\href
  {\doibase 10.1103/PhysRevLett.95.055702} {\bibfield  {journal} {\bibinfo
  {journal} {Physical Review Letters}\ }\textbf {\bibinfo {volume} {95}},\
  \bibinfo {pages} {055702} (\bibinfo {year} {2005})}\BibitemShut {NoStop}%
\bibitem [{\citenamefont {Zhao}\ \emph {et~al.}(2013)\citenamefont {Zhao},
  \citenamefont {Simon},\ and\ \citenamefont {McKenna}}]{zhao2013using}%
  \BibitemOpen
  \bibfield  {author} {\bibinfo {author} {\bibfnamefont {J.}~\bibnamefont
  {Zhao}}, \bibinfo {author} {\bibfnamefont {S.~L.}\ \bibnamefont {Simon}}, \
  and\ \bibinfo {author} {\bibfnamefont {G.~B.}\ \bibnamefont {McKenna}},\
  }\href@noop {} {\bibfield  {journal} {\bibinfo  {journal} {Nature
  Communications}\ }\textbf {\bibinfo {volume} {4}},\ \bibinfo {pages} {1783}
  (\bibinfo {year} {2013})}\BibitemShut {NoStop}%
\bibitem [{\citenamefont {Raty}\ \emph {et~al.}(2015)\citenamefont {Raty},
  \citenamefont {Zhang}, \citenamefont {Luckas}, \citenamefont {Chen},
  \citenamefont {Mazzarello}, \citenamefont {Bichara},\ and\ \citenamefont
  {Wuttig}}]{Raty2015}%
  \BibitemOpen
  \bibfield  {author} {\bibinfo {author} {\bibfnamefont {J.~Y.}\ \bibnamefont
  {Raty}}, \bibinfo {author} {\bibfnamefont {W.}~\bibnamefont {Zhang}},
  \bibinfo {author} {\bibfnamefont {J.}~\bibnamefont {Luckas}}, \bibinfo
  {author} {\bibfnamefont {C.}~\bibnamefont {Chen}}, \bibinfo {author}
  {\bibfnamefont {R.}~\bibnamefont {Mazzarello}}, \bibinfo {author}
  {\bibfnamefont {C.}~\bibnamefont {Bichara}}, \ and\ \bibinfo {author}
  {\bibfnamefont {M.}~\bibnamefont {Wuttig}},\ }\href {\doibase
  10.1038/ncomms8467} {\bibfield  {journal} {\bibinfo  {journal} {Nature
  Communications}\ }\textbf {\bibinfo {volume} {6}},\ \bibinfo {pages} {2041}
  (\bibinfo {year} {2015})}\BibitemShut {NoStop}%
\bibitem [{\citenamefont {Wang}\ \emph {et~al.}(2006)\citenamefont {Wang},
  \citenamefont {Song},\ and\ \citenamefont {Makse}}]{Wang2006}%
  \BibitemOpen
  \bibfield  {author} {\bibinfo {author} {\bibfnamefont {P.}~\bibnamefont
  {Wang}}, \bibinfo {author} {\bibfnamefont {C.}~\bibnamefont {Song}}, \ and\
  \bibinfo {author} {\bibfnamefont {H.~A.}\ \bibnamefont {Makse}},\ }\href
  {\doibase 10.1038/nphys366} {\bibfield  {journal} {\bibinfo  {journal}
  {Nature Physics}\ }\textbf {\bibinfo {volume} {2}},\ \bibinfo {pages} {526}
  (\bibinfo {year} {2006})}\BibitemShut {NoStop}%
\bibitem [{\citenamefont {Odegard}\ and\ \citenamefont
  {Bandyopadhyay}(2011)}]{odegard2011physical}%
  \BibitemOpen
  \bibfield  {author} {\bibinfo {author} {\bibfnamefont {G.}~\bibnamefont
  {Odegard}}\ and\ \bibinfo {author} {\bibfnamefont {A.}~\bibnamefont
  {Bandyopadhyay}},\ }\href@noop {} {\bibfield  {journal} {\bibinfo  {journal}
  {Journal of polymer science Part B: Polymer physics}\ }\textbf {\bibinfo
  {volume} {49}},\ \bibinfo {pages} {1695} (\bibinfo {year}
  {2011})}\BibitemShut {NoStop}%
\bibitem [{\citenamefont {Martin}(1993)}]{martin1993aging}%
  \BibitemOpen
  \bibfield  {author} {\bibinfo {author} {\bibfnamefont {B.}~\bibnamefont
  {Martin}},\ }\href@noop {} {\bibfield  {journal} {\bibinfo  {journal}
  {Calcified Tissue International}\ }\textbf {\bibinfo {volume} {53}},\
  \bibinfo {pages} {S34} (\bibinfo {year} {1993})}\BibitemShut {NoStop}%
\bibitem [{\citenamefont {McKenna}\ \emph {et~al.}(1995)\citenamefont
  {McKenna}, \citenamefont {Leterrier},\ and\ \citenamefont
  {Schultheisz}}]{mckenna1995evolution}%
  \BibitemOpen
  \bibfield  {author} {\bibinfo {author} {\bibfnamefont {G.~B.}\ \bibnamefont
  {McKenna}}, \bibinfo {author} {\bibfnamefont {Y.}~\bibnamefont {Leterrier}},
  \ and\ \bibinfo {author} {\bibfnamefont {C.~R.}\ \bibnamefont
  {Schultheisz}},\ }\href@noop {} {\bibfield  {journal} {\bibinfo  {journal}
  {Polymer Engineering \& Science}\ }\textbf {\bibinfo {volume} {35}},\
  \bibinfo {pages} {403} (\bibinfo {year} {1995})}\BibitemShut {NoStop}%
\bibitem [{\citenamefont {Struik}(1977)}]{struik1977physical}%
  \BibitemOpen
  \bibfield  {author} {\bibinfo {author} {\bibfnamefont {L.~C.~E.}\
  \bibnamefont {Struik}},\ }\href@noop {} {\bibfield  {journal} {\bibinfo
  {journal} {Polymer Engineering \& Science}\ }\textbf {\bibinfo {volume}
  {17}},\ \bibinfo {pages} {165} (\bibinfo {year} {1977})}\BibitemShut
  {NoStop}%
\bibitem [{\citenamefont {Binder}\ and\ \citenamefont
  {Kob}(2011)}]{binder2011glassy}%
  \BibitemOpen
  \bibfield  {author} {\bibinfo {author} {\bibfnamefont {K.}~\bibnamefont
  {Binder}}\ and\ \bibinfo {author} {\bibfnamefont {W.}~\bibnamefont {Kob}},\
  }\href@noop {} {\emph {\bibinfo {title} {Glassy materials and disordered
  solids: An introduction to their statistical mechanics}}}\ (\bibinfo
  {publisher} {World Scientific},\ \bibinfo {year} {2011})\BibitemShut
  {NoStop}%
\bibitem [{\citenamefont {Biroli}\ and\ \citenamefont
  {Garrahan}(2013)}]{biroli2013perspective}%
  \BibitemOpen
  \bibfield  {author} {\bibinfo {author} {\bibfnamefont {G.}~\bibnamefont
  {Biroli}}\ and\ \bibinfo {author} {\bibfnamefont {J.~P.}\ \bibnamefont
  {Garrahan}},\ }\href@noop {} {\bibfield  {journal} {\bibinfo  {journal} {The
  Journal of Chemical Physics}\ }\textbf {\bibinfo {volume} {138}},\ \bibinfo
  {pages} {12A301} (\bibinfo {year} {2013})}\BibitemShut {NoStop}%
\bibitem [{\citenamefont {Debenedetti}\ and\ \citenamefont
  {Stillinger}(2001)}]{debenedetti2001supercooled}%
  \BibitemOpen
  \bibfield  {author} {\bibinfo {author} {\bibfnamefont {P.~G.}\ \bibnamefont
  {Debenedetti}}\ and\ \bibinfo {author} {\bibfnamefont {F.~H.}\ \bibnamefont
  {Stillinger}},\ }\href@noop {} {\bibfield  {journal} {\bibinfo  {journal}
  {Nature}\ }\textbf {\bibinfo {volume} {410}},\ \bibinfo {pages} {259}
  (\bibinfo {year} {2001})}\BibitemShut {NoStop}%
\bibitem [{\citenamefont {Turci}\ \emph {et~al.}(2017)\citenamefont {Turci},
  \citenamefont {Royall},\ and\ \citenamefont {Speck}}]{PhysRevX.7.031028}%
  \BibitemOpen
  \bibfield  {author} {\bibinfo {author} {\bibfnamefont {F.}~\bibnamefont
  {Turci}}, \bibinfo {author} {\bibfnamefont {C.~P.}\ \bibnamefont {Royall}}, \
  and\ \bibinfo {author} {\bibfnamefont {T.}~\bibnamefont {Speck}},\ }\href
  {\doibase 10.1103/PhysRevX.7.031028} {\bibfield  {journal} {\bibinfo
  {journal} {Phys. Rev. X}\ }\textbf {\bibinfo {volume} {7}},\ \bibinfo {pages}
  {031028} (\bibinfo {year} {2017})}\BibitemShut {NoStop}%
\bibitem [{\citenamefont {Kob}\ and\ \citenamefont
  {Barrat}(1997)}]{kob1997aging}%
  \BibitemOpen
  \bibfield  {author} {\bibinfo {author} {\bibfnamefont {W.}~\bibnamefont
  {Kob}}\ and\ \bibinfo {author} {\bibfnamefont {J.~L.}\ \bibnamefont
  {Barrat}},\ }\href@noop {} {\bibfield  {journal} {\bibinfo  {journal}
  {Physical Review Letters}\ }\textbf {\bibinfo {volume} {78}},\ \bibinfo
  {pages} {4581} (\bibinfo {year} {1997})}\BibitemShut {NoStop}%
\bibitem [{\citenamefont {Foffi}\ \emph {et~al.}(2004)\citenamefont {Foffi},
  \citenamefont {Zaccarelli}, \citenamefont {Buldyrev}, \citenamefont
  {Sciortino},\ and\ \citenamefont {Tartaglia}}]{foffi2004aging}%
  \BibitemOpen
  \bibfield  {author} {\bibinfo {author} {\bibfnamefont {G.}~\bibnamefont
  {Foffi}}, \bibinfo {author} {\bibfnamefont {E.}~\bibnamefont {Zaccarelli}},
  \bibinfo {author} {\bibfnamefont {S.}~\bibnamefont {Buldyrev}}, \bibinfo
  {author} {\bibfnamefont {F.}~\bibnamefont {Sciortino}}, \ and\ \bibinfo
  {author} {\bibfnamefont {P.}~\bibnamefont {Tartaglia}},\ }\href@noop {}
  {\bibfield  {journal} {\bibinfo  {journal} {The Journal of Chemical Physics}\
  }\textbf {\bibinfo {volume} {120}},\ \bibinfo {pages} {8824} (\bibinfo {year}
  {2004})}\BibitemShut {NoStop}%
\bibitem [{\citenamefont {Warren}\ and\ \citenamefont
  {Rottler}(2009)}]{Warren_2009}%
  \BibitemOpen
  \bibfield  {author} {\bibinfo {author} {\bibfnamefont {M.}~\bibnamefont
  {Warren}}\ and\ \bibinfo {author} {\bibfnamefont {J.}~\bibnamefont
  {Rottler}},\ }\href {\doibase 10.1209/0295-5075/88/58005} {\bibfield
  {journal} {\bibinfo  {journal} {Europhysics Letters}\ }\textbf {\bibinfo
  {volume} {88}},\ \bibinfo {pages} {58005} (\bibinfo {year}
  {2009})}\BibitemShut {NoStop}%
\bibitem [{\citenamefont {Warren}\ and\ \citenamefont
  {Rottler}(2013)}]{PhysRevLett.110.025501}%
  \BibitemOpen
  \bibfield  {author} {\bibinfo {author} {\bibfnamefont {M.}~\bibnamefont
  {Warren}}\ and\ \bibinfo {author} {\bibfnamefont {J.}~\bibnamefont
  {Rottler}},\ }\href {\doibase 10.1103/PhysRevLett.110.025501} {\bibfield
  {journal} {\bibinfo  {journal} {Phys. Rev. Lett.}\ }\textbf {\bibinfo
  {volume} {110}},\ \bibinfo {pages} {025501} (\bibinfo {year}
  {2013})}\BibitemShut {NoStop}%
\bibitem [{\citenamefont {Hutchinson}(1995)}]{hutchinson1995physical}%
  \BibitemOpen
  \bibfield  {author} {\bibinfo {author} {\bibfnamefont {J.~M.}\ \bibnamefont
  {Hutchinson}},\ }\href@noop {} {\bibfield  {journal} {\bibinfo  {journal}
  {Progress in Polymer Science}\ }\textbf {\bibinfo {volume} {20}},\ \bibinfo
  {pages} {703} (\bibinfo {year} {1995})}\BibitemShut {NoStop}%
\bibitem [{\citenamefont {Barrat}\ and\ \citenamefont
  {Kob}(1999)}]{barrat1999fluctuation}%
  \BibitemOpen
  \bibfield  {author} {\bibinfo {author} {\bibfnamefont {J.~L.}\ \bibnamefont
  {Barrat}}\ and\ \bibinfo {author} {\bibfnamefont {W.}~\bibnamefont {Kob}},\
  }\href@noop {} {\bibfield  {journal} {\bibinfo  {journal} {EPL (Europhysics
  Letters)}\ }\textbf {\bibinfo {volume} {46}},\ \bibinfo {pages} {637}
  (\bibinfo {year} {1999})}\BibitemShut {NoStop}%
\bibitem [{\citenamefont {Kob}\ and\ \citenamefont
  {Barrat}(2000)}]{kob2000fluctuations}%
  \BibitemOpen
  \bibfield  {author} {\bibinfo {author} {\bibfnamefont {W.}~\bibnamefont
  {Kob}}\ and\ \bibinfo {author} {\bibfnamefont {J.~L.}\ \bibnamefont
  {Barrat}},\ }\href@noop {} {\bibfield  {journal} {\bibinfo  {journal} {The
  European Physical Journal B: Condensed Matter and Complex Systems}\ }\textbf
  {\bibinfo {volume} {13}},\ \bibinfo {pages} {319} (\bibinfo {year}
  {2000})}\BibitemShut {NoStop}%
\bibitem [{\citenamefont {Kob}\ \emph {et~al.}(2000)\citenamefont {Kob},
  \citenamefont {Barrat}, \citenamefont {Sciortino},\ and\ \citenamefont
  {Tartaglia}}]{kob2000aging}%
  \BibitemOpen
  \bibfield  {author} {\bibinfo {author} {\bibfnamefont {W.}~\bibnamefont
  {Kob}}, \bibinfo {author} {\bibfnamefont {J.-L.}\ \bibnamefont {Barrat}},
  \bibinfo {author} {\bibfnamefont {F.}~\bibnamefont {Sciortino}}, \ and\
  \bibinfo {author} {\bibfnamefont {P.}~\bibnamefont {Tartaglia}},\ }\href@noop
  {} {\bibfield  {journal} {\bibinfo  {journal} {Journal of Physics: Condensed
  Matter}\ }\textbf {\bibinfo {volume} {12}},\ \bibinfo {pages} {6385}
  (\bibinfo {year} {2000})}\BibitemShut {NoStop}%
\bibitem [{\citenamefont {Kawasaki}\ and\ \citenamefont
  {Tanaka}(2014)}]{PhysRevE.89.062315}%
  \BibitemOpen
  \bibfield  {author} {\bibinfo {author} {\bibfnamefont {T.}~\bibnamefont
  {Kawasaki}}\ and\ \bibinfo {author} {\bibfnamefont {H.}~\bibnamefont
  {Tanaka}},\ }\href {\doibase 10.1103/PhysRevE.89.062315} {\bibfield
  {journal} {\bibinfo  {journal} {Phys. Rev. E}\ }\textbf {\bibinfo {volume}
  {89}},\ \bibinfo {pages} {062315} (\bibinfo {year} {2014})}\BibitemShut
  {NoStop}%
\bibitem [{\citenamefont {Warren}\ and\ \citenamefont
  {Rottler}(2007)}]{PhysRevE.76.031802}%
  \BibitemOpen
  \bibfield  {author} {\bibinfo {author} {\bibfnamefont {M.}~\bibnamefont
  {Warren}}\ and\ \bibinfo {author} {\bibfnamefont {J.}~\bibnamefont
  {Rottler}},\ }\href {\doibase 10.1103/PhysRevE.76.031802} {\bibfield
  {journal} {\bibinfo  {journal} {Phys. Rev. E}\ }\textbf {\bibinfo {volume}
  {76}},\ \bibinfo {pages} {031802} (\bibinfo {year} {2007})}\BibitemShut
  {NoStop}%
\bibitem [{\citenamefont {Waseda}\ and\ \citenamefont
  {Egami}(1979)}]{Waseda1979}%
  \BibitemOpen
  \bibfield  {author} {\bibinfo {author} {\bibfnamefont {Y.}~\bibnamefont
  {Waseda}}\ and\ \bibinfo {author} {\bibfnamefont {T.}~\bibnamefont {Egami}},\
  }\href {\doibase 10.1007/BF00561311} {\bibfield  {journal} {\bibinfo
  {journal} {J. Mater. Sci.}\ }\textbf {\bibinfo {volume} {14}},\ \bibinfo
  {pages} {1249} (\bibinfo {year} {1979})}\BibitemShut {NoStop}%
\bibitem [{\citenamefont {Popescu}(1994)}]{popescu1994structural}%
  \BibitemOpen
  \bibfield  {author} {\bibinfo {author} {\bibfnamefont {M.~A.}\ \bibnamefont
  {Popescu}},\ }\href@noop {} {\bibfield  {journal} {\bibinfo  {journal}
  {Journal of non-crystalline solids}\ }\textbf {\bibinfo {volume} {169}},\
  \bibinfo {pages} {155} (\bibinfo {year} {1994})}\BibitemShut {NoStop}%
\bibitem [{\citenamefont {Fan}\ \emph {et~al.}(2014)\citenamefont {Fan},
  \citenamefont {Iwashita},\ and\ \citenamefont {Egami}}]{PhysRevE.89.062313}%
  \BibitemOpen
  \bibfield  {author} {\bibinfo {author} {\bibfnamefont {Y.}~\bibnamefont
  {Fan}}, \bibinfo {author} {\bibfnamefont {T.}~\bibnamefont {Iwashita}}, \
  and\ \bibinfo {author} {\bibfnamefont {T.}~\bibnamefont {Egami}},\ }\href
  {\doibase 10.1103/PhysRevE.89.062313} {\bibfield  {journal} {\bibinfo
  {journal} {Phys. Rev. E}\ }\textbf {\bibinfo {volume} {89}},\ \bibinfo
  {pages} {062313} (\bibinfo {year} {2014})}\BibitemShut {NoStop}%
\bibitem [{Sup()}]{SupplementalMaterial}%
  \BibitemOpen
  \href@noop {} {\emph {\bibinfo {title} {See Supplemental Material at [URL]
  for more details on the hyperparameters used for the machine-learning model,
  principal component analysis, and Shap analysis for different temperatures
  and the active system.}}}\BibitemShut {Stop}%
\bibitem [{\citenamefont {Cubuk}\ \emph {et~al.}(2015)\citenamefont {Cubuk},
  \citenamefont {Schoenholz}, \citenamefont {Rieser}, \citenamefont {Malone},
  \citenamefont {Rottler}, \citenamefont {Durian}, \citenamefont {Kaxiras},\
  and\ \citenamefont {Liu}}]{PhysRevLett.114.108001}%
  \BibitemOpen
  \bibfield  {author} {\bibinfo {author} {\bibfnamefont {E.~D.}\ \bibnamefont
  {Cubuk}}, \bibinfo {author} {\bibfnamefont {S.~S.}\ \bibnamefont
  {Schoenholz}}, \bibinfo {author} {\bibfnamefont {J.~M.}\ \bibnamefont
  {Rieser}}, \bibinfo {author} {\bibfnamefont {B.~D.}\ \bibnamefont {Malone}},
  \bibinfo {author} {\bibfnamefont {J.}~\bibnamefont {Rottler}}, \bibinfo
  {author} {\bibfnamefont {D.~J.}\ \bibnamefont {Durian}}, \bibinfo {author}
  {\bibfnamefont {E.}~\bibnamefont {Kaxiras}}, \ and\ \bibinfo {author}
  {\bibfnamefont {A.~J.}\ \bibnamefont {Liu}},\ }\href {\doibase
  10.1103/PhysRevLett.114.108001} {\bibfield  {journal} {\bibinfo  {journal}
  {Phys. Rev. Lett.}\ }\textbf {\bibinfo {volume} {114}},\ \bibinfo {pages}
  {108001} (\bibinfo {year} {2015})}\BibitemShut {NoStop}%
\bibitem [{\citenamefont {Schoenholz}\ \emph {et~al.}(2016)\citenamefont
  {Schoenholz}, \citenamefont {Cubuk}, \citenamefont {Sussman}, \citenamefont
  {Kaxiras},\ and\ \citenamefont {Liu}}]{Schoenholz2016}%
  \BibitemOpen
  \bibfield  {author} {\bibinfo {author} {\bibfnamefont {S.~S.}\ \bibnamefont
  {Schoenholz}}, \bibinfo {author} {\bibfnamefont {E.~D.}\ \bibnamefont
  {Cubuk}}, \bibinfo {author} {\bibfnamefont {D.~M.}\ \bibnamefont {Sussman}},
  \bibinfo {author} {\bibfnamefont {E.}~\bibnamefont {Kaxiras}}, \ and\
  \bibinfo {author} {\bibfnamefont {A.~J.}\ \bibnamefont {Liu}},\ }\href@noop
  {} {\bibfield  {journal} {\bibinfo  {journal} {Nature Physics}\ }\textbf
  {\bibinfo {volume} {12}},\ \bibinfo {pages} {469} (\bibinfo {year}
  {2016})}\BibitemShut {NoStop}%
\bibitem [{\citenamefont {Cubuk}\ \emph {et~al.}(2017)\citenamefont {Cubuk},
  \citenamefont {Ivancic}, \citenamefont {Schoenholz}, \citenamefont
  {Strickland}, \citenamefont {Basu}, \citenamefont {Davidson}, \citenamefont
  {Fontaine}, \citenamefont {Hor}, \citenamefont {Huang}, \citenamefont
  {Jiang}, \citenamefont {Keim}, \citenamefont {Koshigan}, \citenamefont
  {Lefever}, \citenamefont {Liu}, \citenamefont {Ma}, \citenamefont
  {Magagnosc}, \citenamefont {Morrow}, \citenamefont {Ortiz}, \citenamefont
  {Rieser}, \citenamefont {Shavit}, \citenamefont {Still}, \citenamefont {Xu},
  \citenamefont {Zhang}, \citenamefont {Nordstrom}, \citenamefont {Arratia},
  \citenamefont {Carpick}, \citenamefont {Durian}, \citenamefont {Fakhraai},
  \citenamefont {Jerolmack}, \citenamefont {Lee}, \citenamefont {Li},
  \citenamefont {Riggleman}, \citenamefont {Turner}, \citenamefont {Yodh},
  \citenamefont {Gianola},\ and\ \citenamefont {Liu}}]{Cubuk2017}%
  \BibitemOpen
  \bibfield  {author} {\bibinfo {author} {\bibfnamefont {E.~D.}\ \bibnamefont
  {Cubuk}}, \bibinfo {author} {\bibfnamefont {R.~J.~S.}\ \bibnamefont
  {Ivancic}}, \bibinfo {author} {\bibfnamefont {S.~S.}\ \bibnamefont
  {Schoenholz}}, \bibinfo {author} {\bibfnamefont {D.~J.}\ \bibnamefont
  {Strickland}}, \bibinfo {author} {\bibfnamefont {A.}~\bibnamefont {Basu}},
  \bibinfo {author} {\bibfnamefont {Z.~S.}\ \bibnamefont {Davidson}}, \bibinfo
  {author} {\bibfnamefont {J.}~\bibnamefont {Fontaine}}, \bibinfo {author}
  {\bibfnamefont {J.~L.}\ \bibnamefont {Hor}}, \bibinfo {author} {\bibfnamefont
  {Y.-R.}\ \bibnamefont {Huang}}, \bibinfo {author} {\bibfnamefont
  {Y.}~\bibnamefont {Jiang}}, \bibinfo {author} {\bibfnamefont {N.~C.}\
  \bibnamefont {Keim}}, \bibinfo {author} {\bibfnamefont {K.~D.}\ \bibnamefont
  {Koshigan}}, \bibinfo {author} {\bibfnamefont {J.~A.}\ \bibnamefont
  {Lefever}}, \bibinfo {author} {\bibfnamefont {T.}~\bibnamefont {Liu}},
  \bibinfo {author} {\bibfnamefont {X.-G.}\ \bibnamefont {Ma}}, \bibinfo
  {author} {\bibfnamefont {D.~J.}\ \bibnamefont {Magagnosc}}, \bibinfo {author}
  {\bibfnamefont {E.}~\bibnamefont {Morrow}}, \bibinfo {author} {\bibfnamefont
  {C.~P.}\ \bibnamefont {Ortiz}}, \bibinfo {author} {\bibfnamefont {J.~M.}\
  \bibnamefont {Rieser}}, \bibinfo {author} {\bibfnamefont {A.}~\bibnamefont
  {Shavit}}, \bibinfo {author} {\bibfnamefont {T.}~\bibnamefont {Still}},
  \bibinfo {author} {\bibfnamefont {Y.}~\bibnamefont {Xu}}, \bibinfo {author}
  {\bibfnamefont {Y.}~\bibnamefont {Zhang}}, \bibinfo {author} {\bibfnamefont
  {K.~N.}\ \bibnamefont {Nordstrom}}, \bibinfo {author} {\bibfnamefont {P.~E.}\
  \bibnamefont {Arratia}}, \bibinfo {author} {\bibfnamefont {R.~W.}\
  \bibnamefont {Carpick}}, \bibinfo {author} {\bibfnamefont {D.~J.}\
  \bibnamefont {Durian}}, \bibinfo {author} {\bibfnamefont {Z.}~\bibnamefont
  {Fakhraai}}, \bibinfo {author} {\bibfnamefont {D.~J.}\ \bibnamefont
  {Jerolmack}}, \bibinfo {author} {\bibfnamefont {D.}~\bibnamefont {Lee}},
  \bibinfo {author} {\bibfnamefont {J.}~\bibnamefont {Li}}, \bibinfo {author}
  {\bibfnamefont {R.}~\bibnamefont {Riggleman}}, \bibinfo {author}
  {\bibfnamefont {K.~T.}\ \bibnamefont {Turner}}, \bibinfo {author}
  {\bibfnamefont {A.~G.}\ \bibnamefont {Yodh}}, \bibinfo {author}
  {\bibfnamefont {D.~S.}\ \bibnamefont {Gianola}}, \ and\ \bibinfo {author}
  {\bibfnamefont {A.~J.}\ \bibnamefont {Liu}},\ }\href {\doibase
  10.1126/science.aai8830} {\bibfield  {journal} {\bibinfo  {journal}
  {Science}\ }\textbf {\bibinfo {volume} {358}},\ \bibinfo {pages} {1033}
  (\bibinfo {year} {2017})},\ \Eprint
  {http://arxiv.org/abs/https://www.science.org/doi/pdf/10.1126/science.aai8830}
  {https://www.science.org/doi/pdf/10.1126/science.aai8830} \BibitemShut
  {NoStop}%
\bibitem [{\citenamefont {Landes}\ \emph {et~al.}(2020)\citenamefont {Landes},
  \citenamefont {Biroli}, \citenamefont {Dauchot}, \citenamefont {Liu},\ and\
  \citenamefont {Reichman}}]{PhysRevE.101.010602}%
  \BibitemOpen
  \bibfield  {author} {\bibinfo {author} {\bibfnamefont {F.~m. c.~P.}\
  \bibnamefont {Landes}}, \bibinfo {author} {\bibfnamefont {G.}~\bibnamefont
  {Biroli}}, \bibinfo {author} {\bibfnamefont {O.}~\bibnamefont {Dauchot}},
  \bibinfo {author} {\bibfnamefont {A.~J.}\ \bibnamefont {Liu}}, \ and\
  \bibinfo {author} {\bibfnamefont {D.~R.}\ \bibnamefont {Reichman}},\ }\href
  {\doibase 10.1103/PhysRevE.101.010602} {\bibfield  {journal} {\bibinfo
  {journal} {Phys. Rev. E}\ }\textbf {\bibinfo {volume} {101}},\ \bibinfo
  {pages} {010602} (\bibinfo {year} {2020})}\BibitemShut {NoStop}%
\bibitem [{\citenamefont {Cubuk}\ \emph {et~al.}(2016)\citenamefont {Cubuk},
  \citenamefont {Schoenholz}, \citenamefont {Kaxiras},\ and\ \citenamefont
  {Liu}}]{CubukJPC2016}%
  \BibitemOpen
  \bibfield  {author} {\bibinfo {author} {\bibfnamefont {E.~D.}\ \bibnamefont
  {Cubuk}}, \bibinfo {author} {\bibfnamefont {S.~S.}\ \bibnamefont
  {Schoenholz}}, \bibinfo {author} {\bibfnamefont {E.}~\bibnamefont {Kaxiras}},
  \ and\ \bibinfo {author} {\bibfnamefont {A.~J.}\ \bibnamefont {Liu}},\ }\href
  {\doibase 10.1021/acs.jpcb.6b02144} {\bibfield  {journal} {\bibinfo
  {journal} {The Journal of Physical Chemistry B}\ }\textbf {\bibinfo {volume}
  {120}},\ \bibinfo {pages} {6139} (\bibinfo {year} {2016})},\ \bibinfo {note}
  {pMID: 27092716},\ \Eprint
  {http://arxiv.org/abs/https://doi.org/10.1021/acs.jpcb.6b02144}
  {https://doi.org/10.1021/acs.jpcb.6b02144} \BibitemShut {NoStop}%
\bibitem [{\citenamefont {Richard}\ \emph {et~al.}(2020)\citenamefont
  {Richard}, \citenamefont {Ozawa}, \citenamefont {Patinet}, \citenamefont
  {Stanifer}, \citenamefont {Shang}, \citenamefont {Ridout}, \citenamefont
  {Xu}, \citenamefont {Zhang}, \citenamefont {Morse}, \citenamefont {Barrat},
  \citenamefont {Berthier}, \citenamefont {Falk}, \citenamefont {Guan},
  \citenamefont {Liu}, \citenamefont {Martens}, \citenamefont {Sastry},
  \citenamefont {Vandembroucq}, \citenamefont {Lerner},\ and\ \citenamefont
  {Manning}}]{PhysRevMaterials.4.113609}%
  \BibitemOpen
  \bibfield  {author} {\bibinfo {author} {\bibfnamefont {D.}~\bibnamefont
  {Richard}}, \bibinfo {author} {\bibfnamefont {M.}~\bibnamefont {Ozawa}},
  \bibinfo {author} {\bibfnamefont {S.}~\bibnamefont {Patinet}}, \bibinfo
  {author} {\bibfnamefont {E.}~\bibnamefont {Stanifer}}, \bibinfo {author}
  {\bibfnamefont {B.}~\bibnamefont {Shang}}, \bibinfo {author} {\bibfnamefont
  {S.~A.}\ \bibnamefont {Ridout}}, \bibinfo {author} {\bibfnamefont
  {B.}~\bibnamefont {Xu}}, \bibinfo {author} {\bibfnamefont {G.}~\bibnamefont
  {Zhang}}, \bibinfo {author} {\bibfnamefont {P.~K.}\ \bibnamefont {Morse}},
  \bibinfo {author} {\bibfnamefont {J.-L.}\ \bibnamefont {Barrat}}, \bibinfo
  {author} {\bibfnamefont {L.}~\bibnamefont {Berthier}}, \bibinfo {author}
  {\bibfnamefont {M.~L.}\ \bibnamefont {Falk}}, \bibinfo {author}
  {\bibfnamefont {P.}~\bibnamefont {Guan}}, \bibinfo {author} {\bibfnamefont
  {A.~J.}\ \bibnamefont {Liu}}, \bibinfo {author} {\bibfnamefont
  {K.}~\bibnamefont {Martens}}, \bibinfo {author} {\bibfnamefont
  {S.}~\bibnamefont {Sastry}}, \bibinfo {author} {\bibfnamefont
  {D.}~\bibnamefont {Vandembroucq}}, \bibinfo {author} {\bibfnamefont
  {E.}~\bibnamefont {Lerner}}, \ and\ \bibinfo {author} {\bibfnamefont {M.~L.}\
  \bibnamefont {Manning}},\ }\href {\doibase 10.1103/PhysRevMaterials.4.113609}
  {\bibfield  {journal} {\bibinfo  {journal} {Phys. Rev. Mater.}\ }\textbf
  {\bibinfo {volume} {4}},\ \bibinfo {pages} {113609} (\bibinfo {year}
  {2020})}\BibitemShut {NoStop}%
\bibitem [{\citenamefont {Oyama}\ \emph {et~al.}(2023)\citenamefont {Oyama},
  \citenamefont {Koyama},\ and\ \citenamefont {Kawasaki}}]{oyama22}%
  \BibitemOpen
  \bibfield  {author} {\bibinfo {author} {\bibfnamefont {N.}~\bibnamefont
  {Oyama}}, \bibinfo {author} {\bibfnamefont {S.}~\bibnamefont {Koyama}}, \
  and\ \bibinfo {author} {\bibfnamefont {T.}~\bibnamefont {Kawasaki}},\ }\href
  {\doibase 10.3389/fphy.2022.1007861} {\bibfield  {journal} {\bibinfo
  {journal} {Frontiers in Physics}\ }\textbf {\bibinfo {volume} {10}} (\bibinfo
  {year} {2023}),\ 10.3389/fphy.2022.1007861}\BibitemShut {NoStop}%
\bibitem [{\citenamefont {Tah}\ \emph {et~al.}(2022)\citenamefont {Tah},
  \citenamefont {Ridout},\ and\ \citenamefont {Liu}}]{tah22}%
  \BibitemOpen
  \bibfield  {author} {\bibinfo {author} {\bibfnamefont {I.}~\bibnamefont
  {Tah}}, \bibinfo {author} {\bibfnamefont {S.~A.}\ \bibnamefont {Ridout}}, \
  and\ \bibinfo {author} {\bibfnamefont {A.~J.}\ \bibnamefont {Liu}},\
  }\href@noop {} {\bibfield  {journal} {\bibinfo  {journal} {The Journal of
  Chemical Physics}\ }\textbf {\bibinfo {volume} {157}},\ \bibinfo {pages}
  {124501} (\bibinfo {year} {2022})}\BibitemShut {NoStop}%
\bibitem [{\citenamefont {Jung}\ \emph {et~al.}(2023)\citenamefont {Jung},
  \citenamefont {Biroli},\ and\ \citenamefont {Berthier}}]{jung22}%
  \BibitemOpen
  \bibfield  {author} {\bibinfo {author} {\bibfnamefont {G.}~\bibnamefont
  {Jung}}, \bibinfo {author} {\bibfnamefont {G.}~\bibnamefont {Biroli}}, \ and\
  \bibinfo {author} {\bibfnamefont {L.}~\bibnamefont {Berthier}},\ }\href
  {\doibase 10.1103/PhysRevLett.130.238202} {\bibfield  {journal} {\bibinfo
  {journal} {Phys. Rev. Lett.}\ }\textbf {\bibinfo {volume} {130}},\ \bibinfo
  {pages} {238202} (\bibinfo {year} {2023})}\BibitemShut {NoStop}%
\bibitem [{\citenamefont {Coslovich}\ \emph {et~al.}(2022)\citenamefont
  {Coslovich}, \citenamefont {Jack},\ and\ \citenamefont
  {Paret}}]{coslovich22}%
  \BibitemOpen
  \bibfield  {author} {\bibinfo {author} {\bibfnamefont {D.}~\bibnamefont
  {Coslovich}}, \bibinfo {author} {\bibfnamefont {R.~L.}\ \bibnamefont {Jack}},
  \ and\ \bibinfo {author} {\bibfnamefont {J.}~\bibnamefont {Paret}},\
  }\href@noop {} {\bibfield  {journal} {\bibinfo  {journal} {The Journal of
  Chemical Physics}\ }\textbf {\bibinfo {volume} {157}},\ \bibinfo {pages}
  {204503} (\bibinfo {year} {2022})}\BibitemShut {NoStop}%
\bibitem [{\citenamefont {Ciarella}\ \emph
  {et~al.}(2023{\natexlab{a}})\citenamefont {Ciarella}, \citenamefont
  {Chiappini}, \citenamefont {Boattini}, \citenamefont {Dijkstra},\ and\
  \citenamefont {Janssen}}]{ciarella22c}%
  \BibitemOpen
  \bibfield  {author} {\bibinfo {author} {\bibfnamefont {S.}~\bibnamefont
  {Ciarella}}, \bibinfo {author} {\bibfnamefont {M.}~\bibnamefont {Chiappini}},
  \bibinfo {author} {\bibfnamefont {E.}~\bibnamefont {Boattini}}, \bibinfo
  {author} {\bibfnamefont {M.}~\bibnamefont {Dijkstra}}, \ and\ \bibinfo
  {author} {\bibfnamefont {L.~M.~C.}\ \bibnamefont {Janssen}},\ }\href@noop {}
  {\bibfield  {journal} {\bibinfo  {journal} {Machine Learning: Science and
  Technology}\ }\textbf {\bibinfo {volume} {4}},\ \bibinfo {pages} {025010}
  (\bibinfo {year} {2023}{\natexlab{a}})}\BibitemShut {NoStop}%
\bibitem [{\citenamefont {Alkemade}\ \emph {et~al.}(2023)\citenamefont
  {Alkemade}, \citenamefont {Smallenburg},\ and\ \citenamefont
  {Filion}}]{Alkemade23}%
  \BibitemOpen
  \bibfield  {author} {\bibinfo {author} {\bibfnamefont {R.~M.}\ \bibnamefont
  {Alkemade}}, \bibinfo {author} {\bibfnamefont {F.}~\bibnamefont
  {Smallenburg}}, \ and\ \bibinfo {author} {\bibfnamefont {L.}~\bibnamefont
  {Filion}},\ }\href {\doibase 10.1063/5.0144822} {\bibfield  {journal}
  {\bibinfo  {journal} {The Journal of Chemical Physics}\ }\textbf {\bibinfo
  {volume} {158}} (\bibinfo {year} {2023}),\ 10.1063/5.0144822},\ \bibinfo
  {note} {134512}\BibitemShut {NoStop}%
\bibitem [{\citenamefont {Janzen}\ \emph {et~al.}(2023)\citenamefont {Janzen},
  \citenamefont {Smeets}, \citenamefont {Debets}, \citenamefont {Luo},
  \citenamefont {Storm}, \citenamefont {Janssen},\ and\ \citenamefont
  {Ciarella}}]{janzen23}%
  \BibitemOpen
  \bibfield  {author} {\bibinfo {author} {\bibfnamefont {G.}~\bibnamefont
  {Janzen}}, \bibinfo {author} {\bibfnamefont {X.~L. J.~A.}\ \bibnamefont
  {Smeets}}, \bibinfo {author} {\bibfnamefont {V.~E.}\ \bibnamefont {Debets}},
  \bibinfo {author} {\bibfnamefont {C.}~\bibnamefont {Luo}}, \bibinfo {author}
  {\bibfnamefont {C.}~\bibnamefont {Storm}}, \bibinfo {author} {\bibfnamefont
  {L.~M.~C.}\ \bibnamefont {Janssen}}, \ and\ \bibinfo {author} {\bibfnamefont
  {S.}~\bibnamefont {Ciarella}},\ }\href {\doibase 10.1209/0295-5075/acdf1b}
  {\bibfield  {journal} {\bibinfo  {journal} {Europhysics Letters}\ }\textbf
  {\bibinfo {volume} {143}},\ \bibinfo {pages} {17004} (\bibinfo {year}
  {2023})}\BibitemShut {NoStop}%
\bibitem [{\citenamefont {Schoenholz}\ \emph {et~al.}(2017)\citenamefont
  {Schoenholz}, \citenamefont {Cubuk}, \citenamefont {Kaxiras},\ and\
  \citenamefont {Liu}}]{Schoenholz263}%
  \BibitemOpen
  \bibfield  {author} {\bibinfo {author} {\bibfnamefont {S.~S.}\ \bibnamefont
  {Schoenholz}}, \bibinfo {author} {\bibfnamefont {E.~D.}\ \bibnamefont
  {Cubuk}}, \bibinfo {author} {\bibfnamefont {E.}~\bibnamefont {Kaxiras}}, \
  and\ \bibinfo {author} {\bibfnamefont {A.~J.}\ \bibnamefont {Liu}},\ }\href
  {\doibase 10.1073/pnas.1610204114} {\bibfield  {journal} {\bibinfo  {journal}
  {Proceedings of the National Academy of Sciences}\ }\textbf {\bibinfo
  {volume} {114}},\ \bibinfo {pages} {263} (\bibinfo {year} {2017})},\ \Eprint
  {http://arxiv.org/abs/https://www.pnas.org/content/114/2/263.full.pdf}
  {https://www.pnas.org/content/114/2/263.full.pdf} \BibitemShut {NoStop}%
\bibitem [{\citenamefont {Ciarella}\ \emph
  {et~al.}(2023{\natexlab{b}})\citenamefont {Ciarella}, \citenamefont
  {Khomenko}, \citenamefont {Berthier}, \citenamefont {Mocanu}, \citenamefont
  {Reichman}, \citenamefont {Scalliet},\ and\ \citenamefont
  {Zamponi}}]{ciarella22tls}%
  \BibitemOpen
  \bibfield  {author} {\bibinfo {author} {\bibfnamefont {S.}~\bibnamefont
  {Ciarella}}, \bibinfo {author} {\bibfnamefont {D.}~\bibnamefont {Khomenko}},
  \bibinfo {author} {\bibfnamefont {L.}~\bibnamefont {Berthier}}, \bibinfo
  {author} {\bibfnamefont {F.~C.}\ \bibnamefont {Mocanu}}, \bibinfo {author}
  {\bibfnamefont {D.~R.}\ \bibnamefont {Reichman}}, \bibinfo {author}
  {\bibfnamefont {C.}~\bibnamefont {Scalliet}}, \ and\ \bibinfo {author}
  {\bibfnamefont {F.}~\bibnamefont {Zamponi}},\ }\href {\doibase
  10.1038/s41467-023-39948-7} {\bibfield  {journal} {\bibinfo  {journal} {Nat.
  Commun.}\ }\textbf {\bibinfo {volume} {14}},\ \bibinfo {pages} {4229}
  (\bibinfo {year} {2023}{\natexlab{b}})}\BibitemShut {NoStop}%
\bibitem [{\citenamefont {Bapst}\ \emph {et~al.}(2020)\citenamefont {Bapst},
  \citenamefont {Keck}, \citenamefont {Grabska-Barwi\'nska}, \citenamefont
  {Donner}, \citenamefont {Cubuk}, \citenamefont {Schoenholz}, \citenamefont
  {Obika}, \citenamefont {Nelson}, \citenamefont {Back}, \citenamefont
  {Hassabis},\ and\ \citenamefont {Kohli}}]{Bapst2020}%
  \BibitemOpen
  \bibfield  {author} {\bibinfo {author} {\bibfnamefont {V.}~\bibnamefont
  {Bapst}}, \bibinfo {author} {\bibfnamefont {T.}~\bibnamefont {Keck}},
  \bibinfo {author} {\bibfnamefont {A.}~\bibnamefont {Grabska-Barwi\'nska}},
  \bibinfo {author} {\bibfnamefont {C.}~\bibnamefont {Donner}}, \bibinfo
  {author} {\bibfnamefont {E.~D.}\ \bibnamefont {Cubuk}}, \bibinfo {author}
  {\bibfnamefont {S.~S.}\ \bibnamefont {Schoenholz}}, \bibinfo {author}
  {\bibfnamefont {A.}~\bibnamefont {Obika}}, \bibinfo {author} {\bibfnamefont
  {A.~W.~R.}\ \bibnamefont {Nelson}}, \bibinfo {author} {\bibfnamefont
  {T.}~\bibnamefont {Back}}, \bibinfo {author} {\bibfnamefont {D.}~\bibnamefont
  {Hassabis}}, \ and\ \bibinfo {author} {\bibfnamefont {P.}~\bibnamefont
  {Kohli}},\ }\href@noop {} {\bibfield  {journal} {\bibinfo  {journal} {Nature
  Physics}\ }\textbf {\bibinfo {volume} {16}},\ \bibinfo {pages} {448}
  (\bibinfo {year} {2020})}\BibitemShut {NoStop}%
\bibitem [{\citenamefont {Boattini}\ \emph {et~al.}(2021)\citenamefont
  {Boattini}, \citenamefont {Smallenburg},\ and\ \citenamefont
  {Filion}}]{Boattini2021}%
  \BibitemOpen
  \bibfield  {author} {\bibinfo {author} {\bibfnamefont {E.}~\bibnamefont
  {Boattini}}, \bibinfo {author} {\bibfnamefont {F.}~\bibnamefont
  {Smallenburg}}, \ and\ \bibinfo {author} {\bibfnamefont {L.}~\bibnamefont
  {Filion}},\ }\href@noop {} {\bibfield  {journal} {\bibinfo  {journal}
  {Physical Review Letters}\ }\textbf {\bibinfo {volume} {127}},\ \bibinfo
  {pages} {88007} (\bibinfo {year} {2021})}\BibitemShut {NoStop}%
\bibitem [{\citenamefont {Bishop}(2006)}]{Bishop2006}%
  \BibitemOpen
  \bibfield  {author} {\bibinfo {author} {\bibfnamefont {C.~M.}\ \bibnamefont
  {Bishop}},\ }\href@noop {} {\emph {\bibinfo {title} {Pattern Recognition and
  Machine Learning (Information Science and Statistics)}}}\ (\bibinfo
  {publisher} {Springer-Verlag},\ \bibinfo {address} {Berlin, Heidelberg},\
  \bibinfo {year} {2006})\BibitemShut {NoStop}%
\bibitem [{\citenamefont {Mahmoud}(2021)}]{mahmoud2019parametric}%
  \BibitemOpen
  \bibfield  {author} {\bibinfo {author} {\bibfnamefont {H.~F.}\ \bibnamefont
  {Mahmoud}},\ }\href@noop {} {\bibfield  {journal} {\bibinfo  {journal}
  {International Journal of Statistics and Probability}\ }\textbf {\bibinfo
  {volume} {10}} (\bibinfo {year} {2021})}\BibitemShut {NoStop}%
\bibitem [{\citenamefont {Tokuda}\ \emph {et~al.}(2020)\citenamefont {Tokuda},
  \citenamefont {Fujisawa}, \citenamefont {Packwood}, \citenamefont
  {Kambayashi},\ and\ \citenamefont {Ueda}}]{Tokuda2020}%
  \BibitemOpen
  \bibfield  {author} {\bibinfo {author} {\bibfnamefont {Y.}~\bibnamefont
  {Tokuda}}, \bibinfo {author} {\bibfnamefont {M.}~\bibnamefont {Fujisawa}},
  \bibinfo {author} {\bibfnamefont {D.~M.}\ \bibnamefont {Packwood}}, \bibinfo
  {author} {\bibfnamefont {M.}~\bibnamefont {Kambayashi}}, \ and\ \bibinfo
  {author} {\bibfnamefont {Y.}~\bibnamefont {Ueda}},\ }\href {\doibase
  10.1063/5.0022451} {\bibfield  {journal} {\bibinfo  {journal} {AIP Advances}\
  }\textbf {\bibinfo {volume} {10}},\ \bibinfo {pages} {105110} (\bibinfo
  {year} {2020})},\ \Eprint
  {http://arxiv.org/abs/https://doi.org/10.1063/5.0022451}
  {https://doi.org/10.1063/5.0022451} \BibitemShut {NoStop}%
\bibitem [{\citenamefont {Lennard-Jones}(1931)}]{Lennard_Jones_1931}%
  \BibitemOpen
  \bibfield  {author} {\bibinfo {author} {\bibfnamefont {J.~E.}\ \bibnamefont
  {Lennard-Jones}},\ }\href {\doibase 10.1088/0959-5309/43/5/301} {\bibfield
  {journal} {\bibinfo  {journal} {Proceedings of the Physical Society}\
  }\textbf {\bibinfo {volume} {43}},\ \bibinfo {pages} {461} (\bibinfo {year}
  {1931})}\BibitemShut {NoStop}%
\bibitem [{\citenamefont {Kob}\ and\ \citenamefont
  {Andersen}(1995)}]{kob1995testing}%
  \BibitemOpen
  \bibfield  {author} {\bibinfo {author} {\bibfnamefont {W.}~\bibnamefont
  {Kob}}\ and\ \bibinfo {author} {\bibfnamefont {H.~C.}\ \bibnamefont
  {Andersen}},\ }\href {\doibase 10.1103/PhysRevE.51.4626} {\bibfield
  {journal} {\bibinfo  {journal} {Physical Review E}\ }\textbf {\bibinfo
  {volume} {51}},\ \bibinfo {pages} {4626} (\bibinfo {year}
  {1995})}\BibitemShut {NoStop}%
\bibitem [{\citenamefont {Plimpton}(1995)}]{PLIMPTON19951}%
  \BibitemOpen
  \bibfield  {author} {\bibinfo {author} {\bibfnamefont {S.}~\bibnamefont
  {Plimpton}},\ }\href {\doibase https://doi.org/10.1006/jcph.1995.1039}
  {\bibfield  {journal} {\bibinfo  {journal} {Journal of Computational
  Physics}\ }\textbf {\bibinfo {volume} {117}},\ \bibinfo {pages} {1} (\bibinfo
  {year} {1995})}\BibitemShut {NoStop}%
\bibitem [{\citenamefont {Kloeden}\ and\ \citenamefont
  {Platen}(2011)}]{kloeden2011numerical}%
  \BibitemOpen
  \bibfield  {author} {\bibinfo {author} {\bibfnamefont {P.}~\bibnamefont
  {Kloeden}}\ and\ \bibinfo {author} {\bibfnamefont {E.}~\bibnamefont
  {Platen}},\ }\href {https://books.google.it/books?id=BCvtssom1CMC} {\emph
  {\bibinfo {title} {Numerical Solution of Stochastic Differential
  Equations}}},\ Stochastic Modelling and Applied Probability\ (\bibinfo
  {publisher} {Springer Berlin Heidelberg},\ \bibinfo {year}
  {2011})\BibitemShut {NoStop}%
\bibitem [{\citenamefont {Romanczuk}\ \emph {et~al.}(2012)\citenamefont
  {Romanczuk}, \citenamefont {B{\"a}r}, \citenamefont {Ebeling}, \citenamefont
  {Lindner},\ and\ \citenamefont {Schimansky-Geier}}]{romanczuk2012active}%
  \BibitemOpen
  \bibfield  {author} {\bibinfo {author} {\bibfnamefont {P.}~\bibnamefont
  {Romanczuk}}, \bibinfo {author} {\bibfnamefont {M.}~\bibnamefont {B{\"a}r}},
  \bibinfo {author} {\bibfnamefont {W.}~\bibnamefont {Ebeling}}, \bibinfo
  {author} {\bibfnamefont {B.}~\bibnamefont {Lindner}}, \ and\ \bibinfo
  {author} {\bibfnamefont {L.}~\bibnamefont {Schimansky-Geier}},\ }\href@noop
  {} {\bibfield  {journal} {\bibinfo  {journal} {The European Physical Journal
  Special Topics}\ }\textbf {\bibinfo {volume} {202}},\ \bibinfo {pages} {1}
  (\bibinfo {year} {2012})}\BibitemShut {NoStop}%
\bibitem [{\citenamefont {Ramaswamy}(2010)}]{ramaswamy2010mechanics}%
  \BibitemOpen
  \bibfield  {author} {\bibinfo {author} {\bibfnamefont {S.}~\bibnamefont
  {Ramaswamy}},\ }\href@noop {} {\bibfield  {journal} {\bibinfo  {journal}
  {Annual Review of Condensed Matter Physics}\ }\textbf {\bibinfo {volume}
  {1}},\ \bibinfo {pages} {323} (\bibinfo {year} {2010})}\BibitemShut {NoStop}%
\bibitem [{\citenamefont {Lindner}\ and\ \citenamefont
  {Nicola}(2008)}]{lindner2008diffusion}%
  \BibitemOpen
  \bibfield  {author} {\bibinfo {author} {\bibfnamefont {B.}~\bibnamefont
  {Lindner}}\ and\ \bibinfo {author} {\bibfnamefont {E.}~\bibnamefont
  {Nicola}},\ }\href@noop {} {\bibfield  {journal} {\bibinfo  {journal} {The
  European Physical Journal Special Topics}\ }\textbf {\bibinfo {volume}
  {157}},\ \bibinfo {pages} {43} (\bibinfo {year} {2008})}\BibitemShut
  {NoStop}%
\bibitem [{\citenamefont {L{\"o}wen}(2020)}]{lowen2020inertial}%
  \BibitemOpen
  \bibfield  {author} {\bibinfo {author} {\bibfnamefont {H.}~\bibnamefont
  {L{\"o}wen}},\ }\href@noop {} {\bibfield  {journal} {\bibinfo  {journal} {The
  Journal of Chemical Physics}\ }\textbf {\bibinfo {volume} {152}},\ \bibinfo
  {pages} {040901} (\bibinfo {year} {2020})}\BibitemShut {NoStop}%
\bibitem [{\citenamefont {Shaebani}\ \emph {et~al.}(2020)\citenamefont
  {Shaebani}, \citenamefont {Wysocki}, \citenamefont {Winkler}, \citenamefont
  {Gompper},\ and\ \citenamefont {Rieger}}]{Shaebani2020}%
  \BibitemOpen
  \bibfield  {author} {\bibinfo {author} {\bibfnamefont {M.~R.}\ \bibnamefont
  {Shaebani}}, \bibinfo {author} {\bibfnamefont {A.}~\bibnamefont {Wysocki}},
  \bibinfo {author} {\bibfnamefont {R.~G.}\ \bibnamefont {Winkler}}, \bibinfo
  {author} {\bibfnamefont {G.}~\bibnamefont {Gompper}}, \ and\ \bibinfo
  {author} {\bibfnamefont {H.}~\bibnamefont {Rieger}},\ }\href {\doibase
  10.1038/s42254-020-0152-1} {\bibfield  {journal} {\bibinfo  {journal} {Nature
  Reviews Physics}\ }\textbf {\bibinfo {volume} {2}},\ \bibinfo {pages} {181}
  (\bibinfo {year} {2020})}\BibitemShut {NoStop}%
\bibitem [{\citenamefont {Dabelow}\ \emph {et~al.}(2019)\citenamefont
  {Dabelow}, \citenamefont {Bo},\ and\ \citenamefont
  {Eichhorn}}]{PhysRevX.9.021009}%
  \BibitemOpen
  \bibfield  {author} {\bibinfo {author} {\bibfnamefont {L.}~\bibnamefont
  {Dabelow}}, \bibinfo {author} {\bibfnamefont {S.}~\bibnamefont {Bo}}, \ and\
  \bibinfo {author} {\bibfnamefont {R.}~\bibnamefont {Eichhorn}},\ }\href
  {\doibase 10.1103/PhysRevX.9.021009} {\bibfield  {journal} {\bibinfo
  {journal} {Phys. Rev. X}\ }\textbf {\bibinfo {volume} {9}},\ \bibinfo {pages}
  {021009} (\bibinfo {year} {2019})}\BibitemShut {NoStop}%
\bibitem [{\citenamefont {Z{\"o}ttl}\ and\ \citenamefont
  {Stark}(2016)}]{zottl2016emergent}%
  \BibitemOpen
  \bibfield  {author} {\bibinfo {author} {\bibfnamefont {A.}~\bibnamefont
  {Z{\"o}ttl}}\ and\ \bibinfo {author} {\bibfnamefont {H.}~\bibnamefont
  {Stark}},\ }\href@noop {} {\bibfield  {journal} {\bibinfo  {journal} {Journal
  of Physics: Condensed Matter}\ }\textbf {\bibinfo {volume} {28}},\ \bibinfo
  {pages} {253001} (\bibinfo {year} {2016})}\BibitemShut {NoStop}%
\bibitem [{\citenamefont {Flenner}\ and\ \citenamefont
  {Szamel}(2015)}]{flenner2015fundamental}%
  \BibitemOpen
  \bibfield  {author} {\bibinfo {author} {\bibfnamefont {E.}~\bibnamefont
  {Flenner}}\ and\ \bibinfo {author} {\bibfnamefont {G.}~\bibnamefont
  {Szamel}},\ }\href@noop {} {\bibfield  {journal} {\bibinfo  {journal} {Nature
  Communications}\ }\textbf {\bibinfo {volume} {6}},\ \bibinfo {pages} {7392}
  (\bibinfo {year} {2015})}\BibitemShut {NoStop}%
\bibitem [{\citenamefont {Li}\ \emph {et~al.}(2016)\citenamefont {Li},
  \citenamefont {Xu},\ and\ \citenamefont {Wittmer}}]{Li_2016}%
  \BibitemOpen
  \bibfield  {author} {\bibinfo {author} {\bibfnamefont {D.}~\bibnamefont
  {Li}}, \bibinfo {author} {\bibfnamefont {H.}~\bibnamefont {Xu}}, \ and\
  \bibinfo {author} {\bibfnamefont {J.~P.}\ \bibnamefont {Wittmer}},\ }\href
  {\doibase 10.1088/0953-8984/28/4/045101} {\bibfield  {journal} {\bibinfo
  {journal} {Journal of Physics: Condensed Matter}\ }\textbf {\bibinfo {volume}
  {28}},\ \bibinfo {pages} {045101} (\bibinfo {year} {2016})}\BibitemShut
  {NoStop}%
\bibitem [{\citenamefont {Janzen}\ and\ \citenamefont
  {Janssen}(2022)}]{janzen2021aging}%
  \BibitemOpen
  \bibfield  {author} {\bibinfo {author} {\bibfnamefont {G.}~\bibnamefont
  {Janzen}}\ and\ \bibinfo {author} {\bibfnamefont {L.~M.~C.}\ \bibnamefont
  {Janssen}},\ }\href {\doibase 10.1103/PhysRevResearch.4.L012038} {\bibfield
  {journal} {\bibinfo  {journal} {Phys. Rev. Research}\ }\textbf {\bibinfo
  {volume} {4}},\ \bibinfo {pages} {L012038} (\bibinfo {year}
  {2022})}\BibitemShut {NoStop}%
\bibitem [{\citenamefont {Gardner}\ and\ \citenamefont
  {Dorling}(1998)}]{GARDNER19982627}%
  \BibitemOpen
  \bibfield  {author} {\bibinfo {author} {\bibfnamefont {M.}~\bibnamefont
  {Gardner}}\ and\ \bibinfo {author} {\bibfnamefont {S.}~\bibnamefont
  {Dorling}},\ }\href {\doibase https://doi.org/10.1016/S1352-2310(97)00447-0}
  {\bibfield  {journal} {\bibinfo  {journal} {Atmospheric Environment}\
  }\textbf {\bibinfo {volume} {32}},\ \bibinfo {pages} {2627} (\bibinfo {year}
  {1998})}\BibitemShut {NoStop}%
\bibitem [{\citenamefont {Pal}\ and\ \citenamefont
  {Mitra}(1992)}]{Multilayer159058}%
  \BibitemOpen
  \bibfield  {author} {\bibinfo {author} {\bibfnamefont {S.}~\bibnamefont
  {Pal}}\ and\ \bibinfo {author} {\bibfnamefont {S.}~\bibnamefont {Mitra}},\
  }\href {\doibase 10.1109/72.159058} {\bibfield  {journal} {\bibinfo
  {journal} {IEEE Transactions on Neural Networks}\ }\textbf {\bibinfo {volume}
  {3}},\ \bibinfo {pages} {683} (\bibinfo {year} {1992})}\BibitemShut {NoStop}%
\bibitem [{\citenamefont {Pedregosa}\ \emph {et~al.}(2011)\citenamefont
  {Pedregosa}, \citenamefont {Varoquaux}, \citenamefont {Gramfort},
  \citenamefont {Michel}, \citenamefont {Thirion}, \citenamefont {Grisel},
  \citenamefont {Blondel}, \citenamefont {Prettenhofer}, \citenamefont {Weiss},
  \citenamefont {Dubourg} \emph {et~al.}}]{pedregosa2011scikit}%
  \BibitemOpen
  \bibfield  {author} {\bibinfo {author} {\bibfnamefont {F.}~\bibnamefont
  {Pedregosa}}, \bibinfo {author} {\bibfnamefont {G.}~\bibnamefont
  {Varoquaux}}, \bibinfo {author} {\bibfnamefont {A.}~\bibnamefont {Gramfort}},
  \bibinfo {author} {\bibfnamefont {V.}~\bibnamefont {Michel}}, \bibinfo
  {author} {\bibfnamefont {B.}~\bibnamefont {Thirion}}, \bibinfo {author}
  {\bibfnamefont {O.}~\bibnamefont {Grisel}}, \bibinfo {author} {\bibfnamefont
  {M.}~\bibnamefont {Blondel}}, \bibinfo {author} {\bibfnamefont
  {P.}~\bibnamefont {Prettenhofer}}, \bibinfo {author} {\bibfnamefont
  {R.}~\bibnamefont {Weiss}}, \bibinfo {author} {\bibfnamefont
  {V.}~\bibnamefont {Dubourg}},  \emph {et~al.},\ }\href@noop {} {\bibfield
  {journal} {\bibinfo  {journal} {the Journal of machine Learning research}\
  }\textbf {\bibinfo {volume} {12}},\ \bibinfo {pages} {2825} (\bibinfo {year}
  {2011})}\BibitemShut {NoStop}%
\bibitem [{\citenamefont {Haykin}(2009)}]{haykin2009neural}%
  \BibitemOpen
  \bibfield  {author} {\bibinfo {author} {\bibfnamefont {S.}~\bibnamefont
  {Haykin}},\ }\href {https://books.google.nl/books?id=KCwWOAAACAAJ} {\emph
  {\bibinfo {title} {Neural Networks and Learning Machines}}},\ Pearson
  International Edition\ (\bibinfo  {publisher} {Pearson},\ \bibinfo {year}
  {2009})\BibitemShut {NoStop}%
\bibitem [{\citenamefont {Kingma}\ and\ \citenamefont
  {Ba}(2014)}]{kingma2014adam}%
  \BibitemOpen
  \bibfield  {author} {\bibinfo {author} {\bibfnamefont {D.~P.}\ \bibnamefont
  {Kingma}}\ and\ \bibinfo {author} {\bibfnamefont {J.}~\bibnamefont {Ba}},\
  }\href@noop {} {\bibfield  {journal} {\bibinfo  {journal} {arXiv preprint
  arXiv:1412.6980}\ } (\bibinfo {year} {2014})}\BibitemShut {NoStop}%
\bibitem [{\citenamefont {Lundberg}\ and\ \citenamefont
  {Lee}(2017)}]{lundberg2017unified}%
  \BibitemOpen
  \bibfield  {author} {\bibinfo {author} {\bibfnamefont {S.~M.}\ \bibnamefont
  {Lundberg}}\ and\ \bibinfo {author} {\bibfnamefont {S.-I.}\ \bibnamefont
  {Lee}},\ }in\ \href@noop {} {\emph {\bibinfo {booktitle} {Proceedings of the
  31st international conference on neural information processing systems}}}\
  (\bibinfo {year} {2017})\ pp.\ \bibinfo {pages} {4768--4777}\BibitemShut
  {NoStop}%
\bibitem [{\citenamefont {Jolliffe}(2002)}]{jolliffe2002principal}%
  \BibitemOpen
  \bibfield  {author} {\bibinfo {author} {\bibfnamefont {I.~T.}\ \bibnamefont
  {Jolliffe}},\ }\href@noop {} {\emph {\bibinfo {title} {Principal component
  analysis for special types of data}}}\ (\bibinfo  {publisher} {Springer},\
  \bibinfo {year} {2002})\BibitemShut {NoStop}%
\bibitem [{\citenamefont {Mandal}\ and\ \citenamefont
  {Sollich}(2020)}]{mandal2020multiple}%
  \BibitemOpen
  \bibfield  {author} {\bibinfo {author} {\bibfnamefont {R.}~\bibnamefont
  {Mandal}}\ and\ \bibinfo {author} {\bibfnamefont {P.}~\bibnamefont
  {Sollich}},\ }\href {\doibase 10.1103/PhysRevLett.125.218001} {\bibfield
  {journal} {\bibinfo  {journal} {Physical Review Letters}\ }\textbf {\bibinfo
  {volume} {125}},\ \bibinfo {pages} {218001} (\bibinfo {year}
  {2020})}\BibitemShut {NoStop}%
\bibitem [{\citenamefont {Bart\'ok}\ \emph {et~al.}(2013)\citenamefont
  {Bart\'ok}, \citenamefont {Kondor},\ and\ \citenamefont
  {Cs\'anyi}}]{bartok13}%
  \BibitemOpen
  \bibfield  {author} {\bibinfo {author} {\bibfnamefont {A.~P.}\ \bibnamefont
  {Bart\'ok}}, \bibinfo {author} {\bibfnamefont {R.}~\bibnamefont {Kondor}}, \
  and\ \bibinfo {author} {\bibfnamefont {G.}~\bibnamefont {Cs\'anyi}},\
  }\href@noop {} {\bibfield  {journal} {\bibinfo  {journal} {Physical Review
  B}\ }\textbf {\bibinfo {volume} {87}},\ \bibinfo {pages} {184115} (\bibinfo
  {year} {2013})}\BibitemShut {NoStop}%
\bibitem [{\citenamefont {De}\ \emph {et~al.}(2016)\citenamefont {De},
  \citenamefont {Bartók}, \citenamefont {Csányi},\ and\ \citenamefont
  {Ceriotti}}]{C6CP00415F}%
  \BibitemOpen
  \bibfield  {author} {\bibinfo {author} {\bibfnamefont {S.}~\bibnamefont
  {De}}, \bibinfo {author} {\bibfnamefont {A.~P.}\ \bibnamefont {Bartók}},
  \bibinfo {author} {\bibfnamefont {G.}~\bibnamefont {Csányi}}, \ and\
  \bibinfo {author} {\bibfnamefont {M.}~\bibnamefont {Ceriotti}},\ }\href
  {\doibase 10.1039/C6CP00415F} {\bibfield  {journal} {\bibinfo  {journal}
  {Phys. Chem. Chem. Phys.}\ }\textbf {\bibinfo {volume} {18}},\ \bibinfo
  {pages} {13754} (\bibinfo {year} {2016})}\BibitemShut {NoStop}%
\end{thebibliography}%
\end{document}